\documentclass[twocolum]{aa}  
\usepackage{amsmath}
\usepackage{textcmds}
\usepackage{graphicx}
\usepackage{xcolor}
\usepackage{txfonts}   
\usepackage{placeins}           
\usepackage{threeparttable} 
\usepackage{orcidlink}
\usepackage{array}
\usepackage{hyperref}

\hypersetup{
      colorlinks,
      citecolor=blue,   
      linkcolor=blue,
      urlcolor=blue
}
\usepackage{ulem}

\newcommand{\be}{\boldsymbol{e}}
\newcommand{\bS}{\boldsymbol{S}}
\newcommand{\bU}{\boldsymbol{U}}
\newcommand{\bJ}{\boldsymbol{J}}
\newcommand{\br}{\boldsymbol{r}}

\newcommand{\buu}{\boldsymbol{u}}

\newcommand{\bnabla}{\boldsymbol{\nabla}}

\newcommand{\inprepcite}[1]{(#1 et al.\ in prep.)}

\definecolor{frcolor}{rgb}{0,0.5,0}

\begin{document} 

\title{Local simulations of common-envelope dynamical inspiral}
\subtitle{Impact of rotation, accretion, and stratification}
\author{
        Damien Gagnier\inst{1,2}\orcidlink{0000-0002-1904-2740}
        \and 
        Giovanni Leidi\inst{2}\orcidlink{0000-0001-7413-7200}
        \and        
        Marco Vetter\inst{1,2}\orcidlink{0009-0007-2322-6001}
        \and 
        Robert Andrassy\inst{1,2}
        \and
         Friedrich K. R{\"o}pke\inst{1,2,3}\orcidlink{0000-0002-4460-0097}
}
\institute{
            Zentrum für Astronomie der Universität Heidelberg, Astronomisches Rechen-Institut, Mönchhofstr. 12-14, D-69120 Heidelberg, Germany\\
            \email{damien.gagnier@uni-heidelberg.de}
        \and
            Heidelberger Institut für Theoretische Studien, Schloss-Wolfsbrunnenweg 35, 69118 Heidelberg, Germany
        \and
            Zentrum für Astronomie der Universität Heidelberg, Institut für Theoretische Astrophysik, Philosophenweg 12, 69120 Heidelberg, Germany
}            
\date{\today}
\abstract{
Common envelope evolution is a crucial phase in binary stellar evolution. Current global three-dimensional simulations lack the resolution to capture the small-scale dynamics around the embedded companion, while local wind-tunnel simulations always approximate the companion's orbital motion as linear rather than as rotation around the center of mass.
We investigated how rotation, accretion, and stratification influence small-scale gas dynamics, gravitational drag and lift forces, and the spin-up rate of the companion.
We performed three-dimensional local hydrodynamic simulations of a $0.2\, M_\odot$ compact companion plunging into the envelope of a $2\, M_\odot$ red giant in a reference frame rotating at the companion's orbital angular velocity, using the \texttt{Athena++} code.
The presence of stratification generates an inward directed force, which is partially opposed by a rotation-induced outward lift force. Both the resulting inward directed force and the drag force, strongly influenced by stratification, would affect the evolution of the binary separation. We propose revised semi-analytical prescriptions for both drag and lift forces. Without accretion and for sufficiently small gravitational softening radii, a quasi-hydrostatic bubble forms around the companion, while accretion prevents its formation and converts kinetic energy into heat that could contribute to the envelope ejection. Drag and lift forces are only marginally affected by accretion. The companion spin-up rate varies non-monotonically in time, first increasing and then decreasing as it plunges deeper into the envelope.
These results motivate future magnetohydrodynamic simulations to investigate how accretion, rotation, and stratification affect magnetic amplification and how magnetic fields, in turn, influence mass and angular momentum accretion rates, as well as the drag and lift force exerted on the companion.}

\maketitle
\nolinenumbers

\section{Introduction}

Common envelope evolution (CEE) is a key phase in the life of many binary star systems. During CEE, a companion star becomes engulfed by the extended envelope of a giant star, which initiates a rapid inspiral that ultimately determines whether the system merges or emerges as a close binary \citep[see, e.g.,][for recent reviews]{Ivanova2020,Roepke2023}. This process is central to the formation of compact binaries that later produce gravitational waves \citep[e.g.,][]{Li2024,Wei2024}, type Ia supernovae \citep[e.g.,][]{Belczynski2005,Ablimit2016,Liu2023}, and other transients. Despite its importance, CEE remains poorly understood, and its simple parameterizations \citep[][]{VDH1976, Tutukov1979, Iben1984, Webbink1984, Livio1988,Nelemans2000,Hirai2022,DiStefano2022} oversimplify the physics and appear inadequate for predicting outcomes from first principles \citep[e.g.,][]{Zorotovic2010,DeMarco2011,Ge2022,Scherbak2023,Torres2025}. 

One of the keys to understanding CEE lies in studying the interaction between a star and the envelope into which it plunges. Current global three-dimensional simulations adequately capture the large-scale evolution of the envelope and the overall interaction between the companion and its surroundings, but they generally lack the resolution to resolve the small-scale flow structure around the embedded object. Local wind-tunnel simulations have been developed to address this limitation \citep[e.g.,][]{MacLeod2015,MM2017,De2020,Prust2024}. By focusing on a small physical region surrounding the companion and imposing an upstream flow, they can achieve much higher spatial resolution. Such models have been used to measure gravitational drag and accretion rates \citep[e.g.,][]{MacLeod2015,MM2017,De2020,Prust2024}, informing semi-analytic prescriptions that complement standard CEE parameterizations in stellar evolution codes \citep[e.g.,][]{Fragos2019,Everson2020,Trani2022}.

Wind-tunnel simulations nevertheless rely on strong simplifying assumptions. They typically neglect rotational and magnetic effects, which may be unavoidable during the dynamical plunge-in phase given the companion's supersonic orbital velocity and the strong amplification of magnetic fields in its vicinity \citep[][]{Ohlmann2016b, Ondratschek2022, Gagnier2024, Vetter2024, Vetter2025}. The embedded object is commonly represented as an absorbing boundary with a radius orders of magnitude larger than its physical size. Such a choice of boundary conditions artificially removes gas before it can form pressure- or rotation-supported structures near the companion and neglects the local dynamics, both of which can affect the flow on larger scales. Wind-tunnel setups also ignore the interaction between the companion and the wake of the primary's core and are thus limited to the dynamical inspiral phase and to relatively low binary mass ratios.

In addition to global and local simulations, efforts have been made to develop analytic and semi-analytic models of the gravitational interaction of one or two massive objects moving through a gaseous medium \citep[e.g.,][]{Ostriker1999, kim2007, kim2008, kim2009, kim2010, Desjacques2022, Bhattacharyya2025}. These models provide useful reference points, though they are formulated under idealized conditions. In reality, the primary's envelope is stratified, turbulence develops in the wake, magnetic fields can be amplified, and the gas dynamics near the object depend sensitively both on how the gravitational potential's singularity is handled and on the sub-grid model for accretion.

In this paper we investigate the gas dynamics in the immediate vicinity of a plunging companion star, with particular emphasis on the roles of rotation, stratification of the primary star's envelope, and accretion. To this end, we solve the equations of hydrodynamics in a reference frame centered on the primary's core and rotating at the companion's orbital angular velocity. We thereby include the Coriolis and centrifugal forces, as well as sink terms for mass, momentum, and energy density.

This paper is organized as follows. In Sect.~\ref{sec:setup} we present our physical model and describe the numerical setup. In Sect.~\ref{sec:results} we show our simulation results and the analysis of the effects of including accretion, rotation, and stratification. In Sect.~\ref{sec:conclusion} we summarize our findings and discuss their implications.

\section{Physical model and numerical setup}\label{sec:setup}
\subsection{Equations of hydrodynamics and the numerical scheme}\label{sec:equations}

We used the \texttt{Athena++} code \citep[][]{Stone2020} to solve the equations of inviscid hydrodynamics in the reference frame centered on the primary's core and rotating at the companion's orbital angular velocity: $ \boldsymbol{\Omega} = \Omega \be_z$
\begin{subequations} \label{eq:full}
\begin{gather}
\frac{\partial \rho}{\partial t} + \bnabla\cdot (\rho \buu) = S_\rho\ , \label{eq:mass} \\[1em]
\begin{split}
\frac{\partial \rho \buu}{\partial t} &+ \bnabla \cdot (\rho \buu \otimes \buu + P \boldsymbol{I}) = - \rho \bnabla \Phi  \\
&\quad - 2 \rho \Omega \left( \be_z \times \buu \right)  - \rho \Omega^2 \be_z  \times \left(\be_z \times \br \right) + \boldsymbol{S}_p \ ,
\end{split} \label{eq:mom} \\[1em]
\begin{split}
\frac{\partial \left(E + \rho \Phi\right)}{\partial t} &+ \bnabla \cdot \left((E+P + \rho \Phi)\buu  \right) =\rho \frac{\partial \Phi}{\partial t}  \  \\
&\quad - \rho \Omega^2 \be_z  \times \left(\be_z \times \br \right) \cdot \buu + S_E \ ,
\end{split} \label{eq:etot} 
\end{gather}
\end{subequations}
where $\mathbf{u}$ is the velocity of a fluid parcel in the rotating frame, $E = e + \rho u^2/2$ with $e$ the volumetric internal energy density, $\Phi$ is the gravitational potential, and we assumed a perfect gas, $P = (\Gamma - 1) e$, with adiabatic index $\Gamma = 5/3$. For simplicity, we fixed the primary's core at $r_1 = 0$ and approximated the center of mass as coinciding with this point, thereby neglecting the linear acceleration of the frame. In this frame, the companion follows a circular orbit in the equatorial plane, and is initially located at $(r, \theta, \varphi) = \{r_2, \pi/2, 0\}$, with $(r, \theta, \varphi)$ corresponding to the conventional spherical coordinates.
The density, momentum, and energy density sink terms, respectively $S_\rho$, $S_p$, and $S_E$, are defined as follows:
\begin{equation}
\begin{aligned}
S_\rho &\coloneqq -\frac{\rho}{\tau_\text{acc}} \mathcal{W}(d, r_s), &
\bS_p &\coloneqq S_\rho \buu^\ast, &
S_E &\coloneqq S_\rho (\epsilon^\ast + \Phi)\ .
\end{aligned}
\end{equation}
Because the companion is unresolved, these terms do not provide realistic accretion rates but do allow us to study the qualitative impact of accretion on the flow. Here, $d = \| \br - \br_2 \| $ is the distance to the companion, $\tau_\text{acc} = (\gamma \Omega)^{-1}$, where $\gamma$ is a dimensionless sink rate.
\begin{equation}\label{eq:kernel}
\mathcal{W}(d, r_s) \coloneqq
\begin{cases}
1 - 6{v}^2 + 6{v}^3, & 0 \le v < \frac{1}{2},\\
2(1-v)^3, & \frac{1}{2} \le v < 1,\\
0, & v \ge 1
\end{cases}
\end{equation}
is a smoothing kernel \citep{Monaghan1985}, where $v= d/r_s$, and $r_s$ is the sink radius. The velocity ($\buu^\ast$) and specific energy ($\epsilon^\ast$) in the frame centered on the companion are given by

\begin{equation}
\buu^\ast \coloneqq \buu \cdot \bigl( \be_r^\ast \be_r^\ast + \delta (\be_\theta^\ast \be_\theta^\ast + \be_\varphi^\ast \be_\varphi^\ast) \bigr), \quad
\epsilon^\ast \coloneqq \frac{1}{2} \buu^\ast \cdot \buu^\ast + \frac{P}{\rho} \ ,
\end{equation}
where
\begin{equation}
    \be_r^\ast \coloneqq \frac{\br - \br_2}{d} \ , \quad  \be_\varphi^\ast \coloneqq \frac{d \left(\be_z \times \be_r^\ast\right) }{\sqrt{d^2 - z^2}} \ , \ \text{and}\  \be_\theta^\ast \coloneqq \be_\varphi^\ast \times \be_r^\ast \ 
\end{equation}
are the spherical unit vectors in the companion's reference frame. Setting $\delta$ to $1$ implies $\buu^\ast = \buu$, which reduces the momentum sink to a standard sink prescription, while $\delta=0$ corresponds to a \qq{torque-free} sink \citep[][]{Dempsey2020,Dittmann2021} where the accretion torque associated with the spin of the sink particle is zero. Choosing $\delta = 0$ is justified because the accreting object is much smaller than the sink radius (see Sect.~\ref{sec:full}). Standard mass removal algorithms, in which the particles accrete a fixed fraction (if not all) of the material entering the sink radius \citep[e.g.,][]{MacLeod2015,MM2017,Munoz2019,Antoni2019,De2020,Tiede2020,Duffell2020,kaaz2023,Rosselli-Calderon2024}, dramatically overestimate such a spin torque. This can generate artificial large-scale angular momentum fluxes, which in turn can significantly affect the flow dynamics on scales much larger than $r_s$ \citep[see][for more details]{Dempsey2020,Dittmann2021}. We mitigated this issue by employing a spline-weighted sink, which progressively reduces the torque contribution from material at increasing distances from the sink.  Finally, we note that, because of angular momentum conservation, removing mass without removing the corresponding angular momentum (i.e., setting $\delta=0$) effectively spins up the gas within the sink radius and in the companion's reference frame. In regimes of strong accretion, this can lead to an artificial increase in the local specific angular momentum, potentially altering the flow dynamics and even resulting in numerical instability. $\Phi(\br,t)$ is the gravitational potential of the binary, 
\begin{equation}\label{eq:fullphi}
    \Phi(\br,t) \coloneqq \sum_{i=1}^2 \Phi_i(\br,t) = -\frac{GM_1}{r} -  GM_2 f(d,h_s) \sigma(t)\ ,
 \end{equation}
where $\sigma(t)$ prevents violent initial transients and ensures numerical stability. It is defined as
\begin{equation}
\sigma(t) \coloneqq \begin{cases} 
    \frac{t}{\tau_{\rm dyn}}, & t < \tau_{\rm dyn} \\ 
    1, & \text{otherwise}  \ .
\end{cases}
\end{equation}
Here, $\tau_{\rm dyn} = \widetilde{R_a}/(r_2\Omega)$  (see Eq.~\ref{eq:tilde_Ra}), $f(d,h)$ is the shape of the softened potential derived from the smoothing kernel of Eq.~\eqref{eq:kernel} \citep[][]{Hernquist1989}:
\begin{align}\label{eq:spline}
  &f(d,h_s) \coloneqq \nonumber\\ &\frac{1}{h_s}\begin{cases}
    \left( -\frac{16}{3}v^{{\prime} 2} + \frac{48}{5}v^{{\prime} 4} -\frac{32}{5}v^{{\prime} 5}  + \frac{14}{5}\right), & \text{if}\ 0 \leq v^{\prime} < \frac{1}{2}.\\[2mm]
     \left( \frac{-1}{15v^{\prime}} - \frac{32}{3}v^{{\prime} 2} +16v^{{\prime} 3} - \frac{48}{5}v^{{\prime} 4} + \frac{32}{15}v^{{\prime} 5} + \frac{16}{5}\right), &  \text{if}\ \frac{1}{2} \leq v^{\prime} < 1.\\[2mm]
   \frac{1}{v^{\prime}}, & \text{if}\ v^{\prime} \ge 1.  
\end{cases}
\end{align}
Here, $v^{\prime} \coloneqq d/h_s$ and $h_s$ is the gravitational softening radius. In this work, we set $r_s = h_s$.

Our simulations are susceptible to numerical shock instabilities when multidimensional shocks align with the grid, often triggering the carbuncle phenomenon, where odd-even perturbations grow unphysically \citep{Quirk1994}. To mitigate this, we employed the low-dissipation Harten-Lax-van Leer Riemann solver \citep[LHLLC;][]{lhll}\footnote{The low-dissipation approach of \cite{lhll} for LHLLD is applied here to HLLC.}, which introduces targeted tangential dissipation to suppress such perturbations. Spatial reconstruction of the primitive variables is performed using the piecewise parabolic method \citep[][]{PPM08}. We disabled the extremum-preserving limiters to improve numerical stability, particularly near the gravitational potential minimum. Time integration uses a strong-stability-preserving third-order Runge-Kutta method \citep[][]{Shu1988,Gottlieb2009}. 

\subsection{Initial conditions}\label{sec:IC}
\subsubsection{Stratified case}\label{sec:stratified_IC}
\begin{figure}
    \centering
    \includegraphics[]{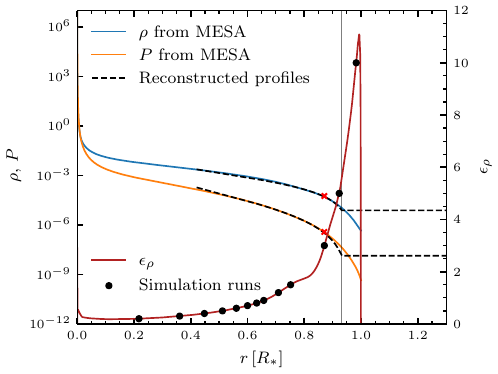}
    \caption{Density and pressure profiles from the $2\,M_\odot$ red giant MESA model of \cite{Ohlmann2016b}, shown as a function of the radius. Dashed lines indicate the polytropic reconstruction used as initial conditions, computed around the location where $\epsilon_\rho = 3$, which is marked by red crosses. The vertical line marks the radius where $\epsilon_\rho = 10$ in the reconstructed stellar structure. The red curve shows the stratification parameter ($\epsilon_\rho$) as a function of radius for the MESA model, while black dots indicate the $\epsilon_\rho$ values employed in our simulations.}
    \label{fig:Prhoini}
\end{figure}

As in \cite{Gagnier2023, Gagnier2024, Gagnier2025}, we initialized the envelope of the primary star in hydrostatic equilibrium with purely radial density and pressure profiles following a polytropic relation with polytropic index $n = 1/(\Gamma - 1)$. The gas obeys an ideal-gas equation of state, and gas self-gravity is neglected.
 Doing so, the local stellar structure around $r = r_2$, for a binary with a mass ratio $q = M_2 /M_1$, can be characterized with two parameters: the density at $r = r_2$ and the ratio between the Hoyle--Lyttleton radius and density scale-height at the same location:
\begin{equation}
    \epsilon_\rho \coloneqq \frac{R_a}{H_\rho} \ ,
\end{equation}
with 
\begin{equation}
R_a \coloneqq \frac{2 G M_2}{u_\infty^2} = \frac{2 q r_2}{1+q} \ , \qquad 
H_\rho \coloneqq -\frac{\rho}{\mathrm{d}\rho / \mathrm{d}r}\Big|_{r=r_2} \ ,
\end{equation}
and where $u_\infty = r_2 \Omega$ is the companion's orbital velocity. We further express the Bondi--Hoyle--Lyttleton radius
 as\begin{equation}\label{eq:tilde_Ra}
\widetilde{R_a} \coloneqq \frac{2GM_2}{u_{\infty}^2 + c_{s,\infty}^2} = \frac{(1+q)^2 \epsilon_\rho }{2q + (1+q)^2 \epsilon_\rho}R_a \ ,
\end{equation} 
where $c_{s, \infty} = c_s(r_2, t=0)$ is the sound speed of the unperturbed envelope at the companion's location. Using the hydrostatic equilibrium equation,
one can derive the density and pressure profiles: 
\begin{align}
  \rho(r) &=  \left[\frac{(1-\Gamma)GM_1}{K\Gamma}  \left( \frac{2q}{(1+q)R_a} \left(1+\frac{2q}{\epsilon_\rho (1+q)(1-\Gamma)} \right) - \frac{1}{r} \right) \right]^n \ , \label{eq:rho} \\ 
  P(r) &= K \rho(r)^\Gamma \ , \label{eq:P}
\end{align}
where
\begin{equation}
    K = \frac{4GM_1q^2}{\rho(r_2)^{1/n}\Gamma(1+q)^2 R_a \epsilon_\rho}\ .
\end{equation}
Equations \eqref{eq:mass}, \eqref{eq:mom}, and \eqref{eq:etot} are homogeneous in density. Varying $\rho(r_2)$ therefore does not change the flow structure or the normalized drag, lift, and accretion coefficients (Sect.~\ref{sec:diag}), but only their absolute values. The $2\,M_\odot$ red giant MESA model shown in Fig.~\ref{fig:Prhoini} illustrates how the stratification parameter ($\epsilon_\rho$) varies across a realistic envelope and provides a reference for assessing the accuracy of our local polytropic reconstruction. The simulation results presented in this paper are expressed in terms of dimensionless parameters and are therefore not tied to this particular stellar model.

In order to minimize the effects of the non-exact numerical hydrostatic equilibrium resulting from the finite grid resolution, we used Gauss-Legendre quadrature of order 12 to map the initial profiles onto the mesh as volume-averaged variables at the barycenter of each cell, which is different from geometric center in polar-spherical coordinates. The initial velocity of the gas in the companion's orbiting frame is set to $\bU = r \sin \theta \Omega \be_\varphi$ at $t=0$, so that the envelope is initially nonrotating in the inertial frame. Crucially, unlike in wind-tunnel simulations, the primary's envelope in our setup is in hydrostatic equilibrium in the absence of the companion.

In our stratified simulations, the domain size is such that it accommodates a sphere of radius $3\, \widetilde{R_a}$, so that the radial extent of the domain is $\Delta r = 6\, \widetilde{R_a}$, and its angular extent is
\begin{equation}
    \Delta \theta = \Delta \varphi \coloneqq 2\arctan{\frac{ 3 \widetilde{R_a}}{\sqrt{r_2^2 - (3 \widetilde{R_a})^2 }}}  \ ,
\end{equation}
where we have set $\varphi_2 = 0$ and $\theta_2 = \pi/2$. We chose a mass ratio $q = 0.1$ so that $\widetilde{R_a} \ll r_2$ and the domain does not include the primary's core.
 In regimes of strong stratification ($\epsilon_\rho \gtrsim 1$), the outer radial boundary of the computational domain can extend beyond the stellar surface, requiring the modeling of an outer vacuum-like atmosphere \citep[e.g.,][]{MacLeod2018, Gagnier2023}. To handle this, we defined a critical radius ($r_c$) at which the transition from the stellar envelope to the atmosphere occurs. In this work, we set $r_c$ to be the radius at which the local stratification parameter of the reconstructed profile reaches a value of 10, coinciding approximately with both the maximum of $ \epsilon_\rho $ and the location of the stellar surface in the one-dimensional MESA model (see Fig.~\ref{fig:Prhoini}). The atmosphere is assumed to be isothermal and in hydrostatic equilibrium, and it is smoothly matched to the envelope by ensuring continuity of the density, pressure, and pressure gradient at $r = r_c$. The resulting atmospheric density and pressure profiles are given by
\begin{align}
        \rho(r \ge r_c) &= \rho(r_c) \exp \left[ \frac{\Gamma GM_1}{(c^a_s)^2} \left(\frac{1}{r} - \frac{1}{r_c} \right) \right] \ , \\
        P(r \ge r_c) &=\frac{\rho (c^a_s)^2}{\Gamma} \ , 
\end{align}
where the atmosphere's sound speed is defined as $c^a_s = \sqrt{K \Gamma \rho(r_c)^{1/n}}$. 

\subsubsection{Non-stratified case}
For the non-stratified simulations presented in Sects. \ref{sec:fiducial}, \ref{sec:accr}, and \ref{sec:rot}, we set $M_2 = 1$, $\Phi_1 = 0$, $R_a = 1$, $\rho(t=0) = 1$, and thus $u_\infty = \sqrt{2}$. The initial gas pressure is set by choosing the Mach number, $P(t=0) = u_\infty^2/(\mathcal{M}_\infty^2 \Gamma)$. The companion's position with respect to the center of rotation is $r_2 = {\rm Ro}\, R_a$ where the Rossby number ${\rm Ro}$ is prescribed, and the orbital angular velocity therefore reads $\Omega = u_\infty/r_2$. In the nonrotating wind-tunnel simulations, we set the wind velocity $\boldsymbol{U} = u_\infty$.

\subsection{Mesh structure and boundary conditions}\label{sec:mesh}
Our root-level grid resolution is $256^3$, on top of which we added four levels of adaptive mesh refinement. We adopted a mesh refinement criterion based on the second derivative error norm of a function $\sigma$ \citep{Lohner1987}. A mesh block is tagged for refinement whenever the following condition is met for a given number of successive cycles \citep[e.g.,][]{PLUTO2012, Gagnier2023}:
\begin{equation}\label{eq:Lohner}
 \chi \coloneqq \frac{\sum_d |\Delta_{d,+1/2} \sigma - \Delta_{d,-1/2} \sigma|^2}{\sum_d \left( |\Delta_{d,+1/2} \sigma| + |\Delta_{d,-1/2} \sigma| + \epsilon \sigma_{d,\rm ref} \right)^2} \ge \chi_r^\prime \ ,
\end{equation}
where, for instance, $\Delta_{r,\pm 1/2} \sigma = \pm (\sigma_{i \pm 1} - \sigma_i)$ and $\sigma_{r,\rm ref} = |\sigma_{i+1}| + 2|\sigma_i| + |\sigma_{i-1}|$. The parameter $\epsilon$ is introduced as a filter to prevent refinement in regions with small-scale fluctuations or noise, and we typically set $\epsilon = 0.01$. In our simulations, we used $\sigma = \mathcal{M} = |\buu| / c_s$ and $\chi_r' = \chi_r + b(1 - \chi_r)$, where $\chi_r = 0.95$. Here, $b$ is defined as the minimum of 1 and the ratio of the total number of $16^3$-cell mesh blocks to a user-specified maximum, which we set to $10^5$. In quasi-steady state, each simulation contains approximately $4.1\times10^8$ cells. A mesh block is tagged for de-refinement whenever $\chi < \chi_d - b(1 - \chi_d)$, with $\chi_d = 0.8$, for 20 successive cycles.
We imposed no-inflow boundary conditions at both radial boundaries and at the minimum-$\varphi$ boundary, and outflow boundary conditions at both $\theta$ boundaries. Pressure and density in the ghost cells at the inner radial boundary and at the maximum-$\varphi$ boundary follow Eqs.~(\ref{eq:rho}) and (\ref{eq:P}), with the azimuthal velocity in the ghost cells at the maximum-$\varphi$ boundary set to $\bU$. Quantities in the ghost cells at the maximum-$\varphi$ boundary are evaluated using Gauss-Legendre quadrature.

\subsection{Diagnostics}\label{sec:diag}
In this section we summarize the diagnostics computed at runtime in our simulations.

\subsubsection{Accreted quantities}
The total mass accretion rate onto the sink particle is
\begin{equation}
    \dot{M} \coloneqq -\int S_\rho \, \mathrm{d}V \ .
\end{equation}
The spin torque, associated with the change in the sink's internal angular momentum, is
\begin{equation}
  \dot{\bJ}_{\rm acc}^{\rm spin} \coloneqq - \delta \int S_\rho \, (\br - \br_2) \times \buu \cdot (\be_\theta^\ast \be_\theta^\ast + \be_\varphi^\ast \be_\varphi^\ast) \, \mathrm{d}V \ .
\end{equation}
Here, $\delta = 0$ corresponds to neglecting the spin torque.  
The linear momentum transfer rate in the direction $i$ due to accretion is
\begin{equation}
 \dot{p}_i \coloneqq  -\int S_\rho \, (\buu^\ast \cdot \be_i) \, \mathrm{d}V \ .
\end{equation}

\subsubsection{Drag and lift forces}

The drag force, integrated within a sphere of radius $R_i$ of volume $V_i$ around the companion, is
\begin{equation}\label{eq:fdrag}
\begin{aligned}
F_{\rm drag}(V_i) &\coloneqq -\int_{V_i} \rho \, \nabla \Phi_2 \cdot \frac{\bU}{\|\bU\|} \, \mathrm{d}V, 
\quad V_i = \left\{ \|\mathbf{r} - \mathbf{r}_2\| \leq R_i \right\} \\
&= \int_{V_i} \frac{\rho\, GM_2\, r_2 \sin \varphi}{h^2 u^{\prime}} \frac{\partial f(u^{\prime})}{\partial u^{\prime}} \, \mathrm{d}V \ .
\end{aligned}
\end{equation}
Similarly, the radial (lift) force, integrated within the same volume, is
\begin{equation}\label{eq:fperp}
\begin{aligned}
F_\perp(V_i) &\coloneqq \int_{V_i} \rho \, \boldsymbol{\nabla} \Phi_2 \cdot \frac{\boldsymbol{\nabla} \Phi_1}{\| \boldsymbol{\nabla} \Phi_1 \|} \, \mathrm{d}V,
\quad V_i = \left\{ \| \mathbf{r} - \mathbf{r}_2 \| \leq R_i \right\} \\
&= -\int_{V_i} \frac{\rho\, GM_2 \left(r - r_2 \sin \theta \cos \varphi \right)}{h^2 u^{\prime}} \frac{\partial f(u^{\prime})}{\partial u^{\prime}} \, \mathrm{d}V \ .
\end{aligned}
\end{equation}
In this work, we neglected the hydrodynamic drag and lift forces, as they are expected to be orders of magnitude weaker than their gravitational counterparts \citep[e.g.,][]{Ricker2012}, with $
F_{\rm hydro}/F_{\rm drag} \sim (R_2/R_a)^2 \ll 1$,
where $R_2$ is the radius of the companion.

\section{Results}\label{sec:results}
\subsection{Nonrotating, non-stratified, and non-accreting case}\label{sec:fiducial}

\begin{figure}
    \centering
    \includegraphics[]{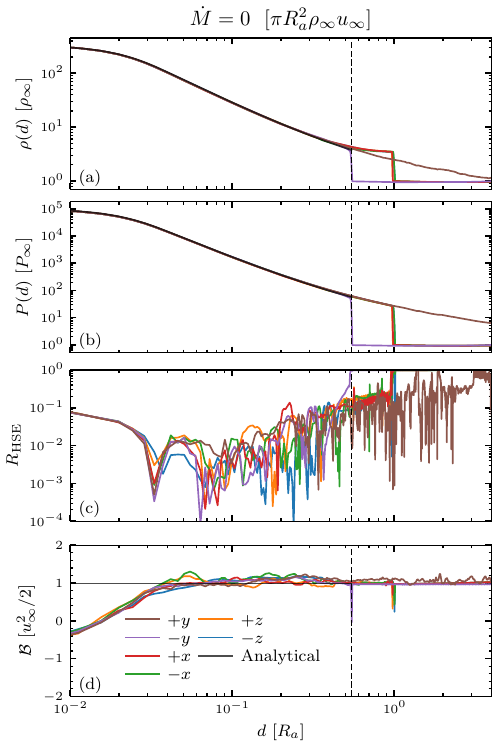}
 \caption{
Density (panel a), pressure (panel b), normalized hydrostatic balance residual $
R_{\rm HSE} = \left\lVert \nabla P + \rho \, \nabla \Phi_2 \right\rVert/\max \left( \lVert \nabla P \rVert, \; \lVert \rho \, \nabla \Phi_2 \rVert \right)
$  (panel c), and Bernoulli parameter  (panel d) profiles along radial rays from the companion for a nonrotating, non-stratified, and non-accreting simulation with $\mathcal{M}_\infty = 4$. Gray lines indicate the analytical solutions for density and pressure (Eqs.~\ref{eq:rho_HSE}, \ref{eq:P_HSE}) and the hydrostatic Bernoulli parameter $\mathcal{B}_{\rm HSE}$ (Eq.~\ref{eq:BHSE}). The vertical dashed line marks the predicted shock radius, $d_s$ (Eq.~\ref{eq:Rs}). All quantities are averaged over a duration of $\Delta t = R_a / u_\infty$ once a quasi-steady state has been reached.}
    \label{fig:HSE_bubble}
\end{figure}

As a fiducial model, we considered wind-tunnel simulations in which a uniform supersonic flow impinges on a non-accreting point mass, similar to the setups of \citet{Thun2016} and \citet{Prust2024}. In our case, however, the companion is neither a solid sphere nor an absorbing boundary, but instead is represented by a softened gravitational potential. This is the same treatment used for the companion and the primary’s core in global common-envelope simulations. Our local, high-resolution models allow us to resolve the small-scale flow around the companion. For a companion star of radius $R_2$, the degree of flow nonlinearity can be quantified by the parameter \citep{kim2009}
\begin{equation}
    \eta \coloneqq \frac{1}{2} \frac{\mathcal{M}_\infty^2}{\mathcal{M}_\infty^2-1} \frac{R_a}{R_2} \ ,
\end{equation}
where $\mathcal{M}_\infty = \mathcal{M}(r_2, t=0)$ is the Mach number of the companion in the unperturbed envelope. \citet{Prust2024} show that when $\eta \gg 1$, a hydrostatic halo forms around the companion. In our setup, $R_2 \to 0$ and thus $\eta \to \infty$. We therefore expect the formation of a hydrostatic bubble, analogous to the halos reported in \citet{Thun2016} and \citet{Prust2024}.  Assuming such a bubble exists and extends up to the shock front, its density and pressure profiles follow directly from hydrostatic equilibrium: 
\begin{align}
    \rho(d) &= \rho(d_s) \left[ 1 - \frac{\Gamma - 1} {c_{s, \rm ps}^2} \left( \Phi_2 - \Phi_2(d_s) \right) \right ]^n \ , \label{eq:rho_HSE} \\ 
    P(d) &= K_2 \rho(d)^\Gamma  \label{eq:P_HSE}\ .
\end{align}
Here 
\begin{equation}\label{eq:Rs}
    d_s \coloneqq \kappa \frac{R_a}{2}\frac{\mathcal{M}_\infty^2}{\mathcal{M}_\infty^2 - 1} \ 
\end{equation}
is the shock stand-off distance, with $\kappa$ a factor of order unity \citep[][]{Thun2016}, which we set to 1, and  
\begin{equation}
    K_2 =  \frac{P_\infty}{\rho_\infty^\Gamma} 
 \frac{\left[ (\Gamma - 1)(3 \mathcal{M}_\infty^2 - 2) + 2 \right]^\Gamma \left[ (\Gamma + 1) + 6 \Gamma (\mathcal{M}_\infty^2 - 1) \right] }{ \left[ (\Gamma + 1)(3 \mathcal{M}_\infty^2 - 2) \right]^\Gamma }
\ ,
\end{equation}
where $\rho_\infty$ and $P_\infty$ are respectively the density and pressure of the unperturbed envelope at the companion's location. Using the Rankine--Hugoniot conditions together with pre-shock energy conservation, $u_\infty^2/2 = u_{\rm ps}^2/2 - GM/d_s$, the post-shock sound speed is  ${c_{s, \rm ps}^2} = {c_{s, \infty}^2}\, \xi(\mathcal{M}_\infty,\Gamma)$, where
\begin{equation}
    \xi(\mathcal{M}_\infty,\Gamma) = 
\frac{\Big( 2\Gamma \big(3\mathcal{M}_\infty^2 - 2\big) - (\Gamma - 1) \Big)\,
\Big( (\Gamma - 1)\big(3\mathcal{M}_\infty^2 - 2\big) + 2 \Big)}
{(\Gamma + 1)^2 \,\big(3\mathcal{M}_\infty^2 - 2\big)}\ .
\end{equation}
Figure~\ref{fig:HSE_bubble} compares these analytical density and pressure profiles with results from a nonrotating, non-stratified, and non-accreting wind-tunnel simulation with $\mathcal{M}_\infty \coloneqq u_\infty/c_{s,\infty} = 4$, and $h_s = 0.05\, R_a$. Profiles are shown along the six Cartesian radial directions centered on the companion ($\pm x$, $\pm y$, $\pm z$), within the $xy$ and $xz$ planes. The comparison shows excellent agreement: both the analytical profiles and the predicted shock location from Eq.~\eqref{eq:Rs} reproduce the simulation results with high accuracy. This indicates that a quasi-hydrostatic bubble indeed forms around the point mass and extends out to the shock front.  
To assess whether this bubble is energetically bound or unbound, we evaluated the Bernoulli parameter:  
\begin{equation}
    \mathcal{B}(\br) \coloneqq \frac{1}{2} \|\buu\|^2 + h(\br) + \Phi_2(\br) \ . 
\end{equation}
where $h=(P+e)/\rho$ is the specific enthalpy. By definition, within the quasi-hydrostatic bubble, $\|\buu\| \simeq 0$ and $\partial_d \mathcal{B} \simeq 0$, so  
\begin{equation}
\begin{aligned}
    \mathcal{B}(d \leq d_s )  \simeq \mathcal{B}_{\rm HSE} &\coloneqq h(d_s) + \Phi_2(d_s) \\
     & = {c_{s, \infty}^2} \left( \frac{\xi(\mathcal{M}_\infty,\Gamma)}{\Gamma - 1} - \mathcal{M}_\infty^2 + 1 \right)\ . \label{eq:BHSE}
    \end{aligned}
    \end{equation}
Thus, the condition for the gas bubble to be formally gravitationally bound ($\mathcal{B}_{\rm HSE} \leq 0$) reduces to a simple inequality involving only $\Gamma$ and $\mathcal{M}_\infty$:
\begin{equation}\label{eq:cond}
     \left( \frac{\xi(\mathcal{M}_\infty,\Gamma)}{\Gamma - 1} - \mathcal{M}_\infty + 1 \right) \leq 0 \ .
\end{equation}
For adiabatic indices relevant to stellar interiors($1<\Gamma<5/3$), condition~\eqref{eq:cond} is never satisfied for supersonic flows ($\mathcal{M}_\infty \geq 1$).  
In other words, no formally gravitationally bound hydrostatic bubble can form around a point mass in the standard wind-tunnel setup. The gas is energetically unbound, so local perturbations at the bubble's surface, such as turbulence or small density fluctuations, could in principle trigger expansion and allow some material to escape. In practice, however, the shock upstream and the continuous inflow downstream act to confine the gas, maintaining a quasi-stable hydrostatic bubble around the point mass.
We verified the theoretical Bernoulli parameter in the post-shock region \eqref{eq:BHSE} as well as the hydrostatic balance residual in Fig.~\ref{fig:HSE_bubble}. Good agreement is found except very close to the point mass, where $h + \Phi_2 \ll |h| \sim |\Phi_2|$. In this region, even small errors $\delta h$ arising from finite resolution or the non-spherically symmetric grid around the companion lead to large fractional deviations in the Bernoulli parameter, $\delta \mathcal{B} / |\mathcal{B}| = \delta h / |\mathcal{B}| \gg 1$.

In a uniform gaseous medium, the gravitational drag exerted on a supersonically moving point mass is expected to grow logarithmically with the integration radius \citep[][]{Ruderman1971},
\begin{equation}\label{eq:Coulomb}
F_{\rm drag} \propto \ln \left(\frac{b_{\rm max}}{b_{\rm min}}\right),
\end{equation} 
where $b_{\rm max}$ is the radius of integration, and $b_{\rm min}$ is the effective inner cutoff for which the linear theory underlying the Coulomb-logarithm scaling Eq.~\eqref{eq:Coulomb} remains valid. 
What sets this cutoff has been discussed extensively in the literature. Often, either the physical radius ($R$) of the object or (for a compact and massive body) $b_{\rm min} \simeq \max(R, R_a)$ is chosen.
\citet{Thun2016} identify $b_{\rm min}$ with the shock standoff distance $d_s$ (Eq.~\ref{eq:Rs}), effectively treating it as the minimum spatial scale of interaction below which gas exerts no drag. 
This interpretation is misleading as $b_{\rm min}$ is not the scale at which the drag vanishes, but rather the smallest radius where the linear Coulomb-logarithm scaling applies. 
Gas inside $b_{\rm min}$ still exerts force, but its contribution does not follow the logarithmic scaling. 
In Fig.~\ref{fig:coulomb} we show the time evolution of the gravitational drag measured within different logarithmically spaced integration radii, scaled by the Coulomb logarithm. 
In a quasi-steady state, the curves are consistent with the Coulomb-logarithm scaling for an effective inner cutoff of $b_{\rm min} \simeq 0.78\,R_a$, about 50\% larger than the shock standoff distance $d_s$. 
In an infinite uniform medium, the drag force would formally diverge, but in practice $b_{\rm max}$ is limited by the causal extent of the wake, $u_\infty t$. 
At $t = 200\,R_a/u_\infty$, this corresponds to $b_{\rm max} \simeq 200\,R_a$, implying that the gravitational drag enclosed within our numerical domain represents only $\sim 24\%$ of the total drag force at that time. This may become a significant limitation for local simulations or global simulations with a limited numerical domain size. Indeed, including stratification, magnetic fields, or accretion may cause the drag force to deviate from a simple Coulomb-logarithm scaling. In such cases, the limited domain may prevent the development of reliable drag force prescriptions (see also Sect.~\ref{sec:draglift}).

\subsection{Accretion effects}\label{sec:accr}
While jet-like outflows are observed in hydrodynamical simulations of CEE \citep[][]{Gagnier2025, Lau2025}, the high-momentum outflows observed in protoplanetary nebulae and planetary nebulae likely require additional energy transfer such as magnetic energy conversion via Lorentz force \citep[][]{Ondratschek2022, Gagnier2024, Vetter2024, Vetter2025} or accretion onto the companion \citep[][]{Ricker2012,Blackman2014, Chamandy2018}.
To investigate the latter, we used the simplified wind-tunnel setup described above but allowed the companion to accrete mass, momentum, and energy.

\begin{figure*}
    \centering
    \includegraphics[]{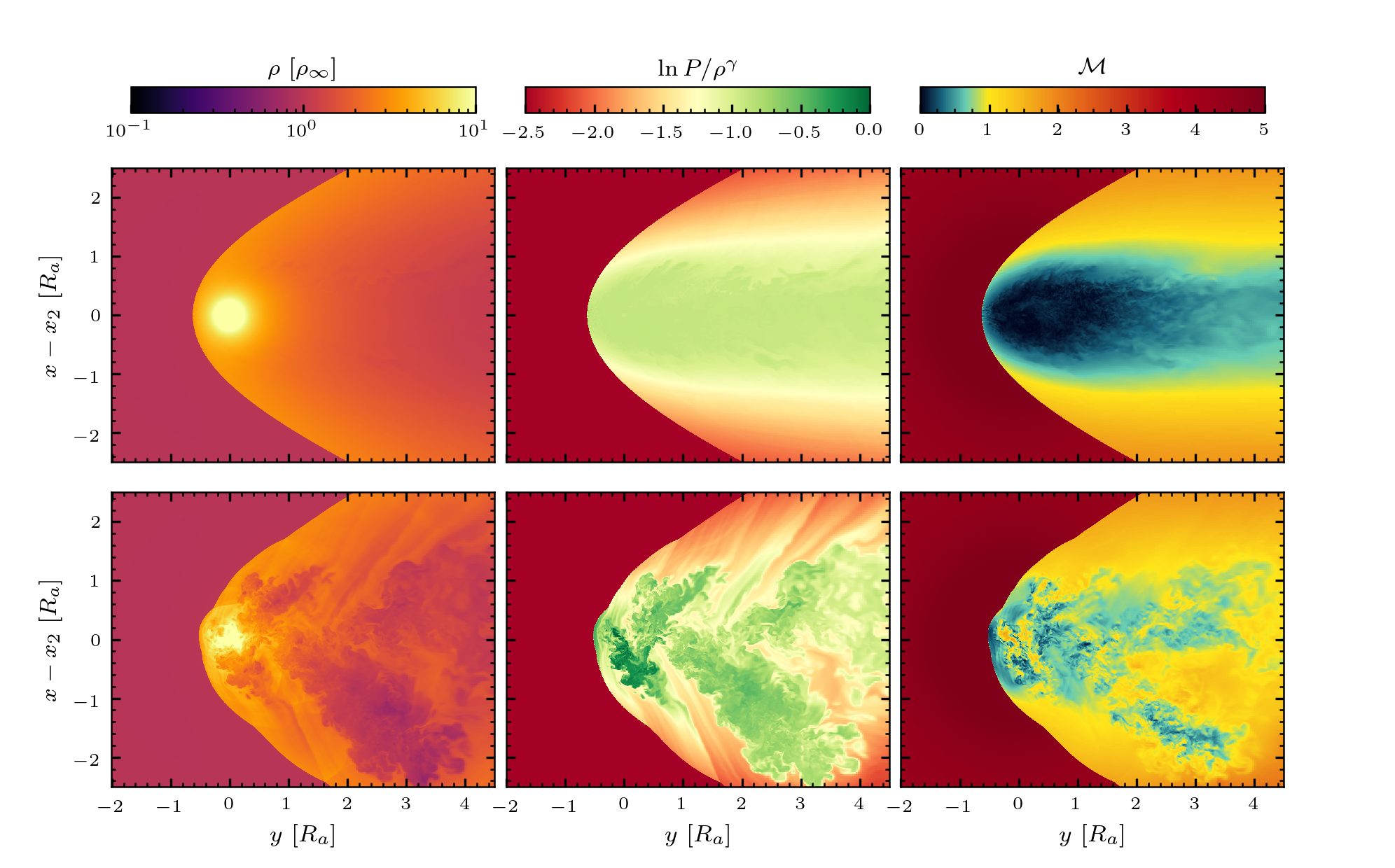}
    \caption{Zoomed-in snapshots in the $xy$ plane of the density, pseudo-entropy, and Mach number at $t = 100\, R_a/u_\infty$ for non-stratified simulations with $h_s = 0.05\, R_a$ and $\mathcal{M}_\infty = 4$. First row: Non-accreting case. Second row: Accreting case with $\gamma = 100$ and $\delta = 0$.}
        \label{fig:snap_acc}
\end{figure*}

In Fig.~\ref{fig:HSE_bubble_acc} we show density, pressure,  normalized hydrostatic balance residual, and Bernoulli parameter profiles along radial rays from the companion for a simulation run with $h_s = 0.05\, R_a$, $\mathcal{M}_\infty = 4$, $\delta = 0$, and $\gamma = 100$. In a quasi-steady state, this corresponds to an accretion rate $\dot{M} \simeq  0.02\, \dot{M}_{\rm HL}$, where $\dot{M}_{\rm HL} = \pi R_a^2 \,\rho_\infty u_\infty$ is the Hoyle--Lyttleton accretion rate \citep{HL1939}. Although such an accretion rate is far below the Hoyle--Lyttleton prediction, its associated inflow is sufficient to destabilize the gas near the companion. Instead of accumulating into a quasi-hydrostatic, pressure-supported bubble, material is continuously drained and funneled through the asymmetric wake. This sustained inflow creates strong velocity shear near the companion, driving turbulence and inducing collisions between converging streams. Accretion contributes to local vorticity through the sink term $-(1/\tau_{\rm acc})\bnabla \times (\mathcal{W} \mathbf{u}^\ast)$, which vanishes only when $\mathbf{u}^\ast = \mathbf{u}_d(d)$. Additional vorticity is generated by departures from hydrostatic equilibrium and accretion anisotropies (mostly resulting from rotation and/or stratification; see Sects.~\ref{sec:rot} and \ref{sec:strat}), via the baroclinic torque, $(\bnabla \rho \times \bnabla P)/\rho^2$. Collisions between converging streams then produce weak shocks and small-scale viscous dissipation that locally increase the entropy\footnote{With our sink prescription, the entropy equation takes the form
$\rho T \mathrm{D}s/\mathrm{D}t = S_\rho (2-\Gamma) P / ((\Gamma-1) \rho) \leq 0$,
implying that the sink acts as an explicit entropy sink. The entropy generation is therefore not caused by the sink itself, but arises from shocks and from the conversion of turbulent kinetic energy into heat via numerical dissipation.}. The higher-entropy gas is advected downstream, where it expands and forms bubble-like structures in the wake, while simultaneously perturbing the upstream flow and distorting the bow shock, causing the standoff distance to fluctuate. This is illustrated in Fig.~\ref{fig:snap_acc}, where we show density, pseudo-entropy, and Mach number snapshots taken at $t = 100\, R_a/u_\infty$, for the two aforementioned simulations.

\begin{figure}
    \centering
    \includegraphics[]{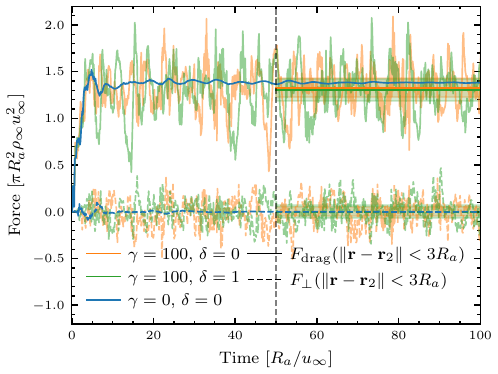}
 \caption{Time evolution of the radial and drag forces exerted by the gas on the companion in non-stratified simulations with ${\rm Ro}= 5.5$ and $\mathcal{M}_\infty = 4$, with and without accretion. Forces are integrated within a sphere of radius $3\, R_a$. The shaded regions indicate the $3\sigma$ range (see Appendix \ref{app:stat}).}
    \label{fig:Drag_acc}
\end{figure}

Such turbulent, anisotropic accretion-driven flows may efficiently amplify magnetic fields through stretching, twisting, and folding of field lines, potentially producing much faster magnetic energy growth near the companion than in the non-accreting regime. Even without accretion, magnetic fields can significantly influence the binary separation evolution, drive jet-like outflows \citep{Ondratschek2022, Vetter2024, Vetter2025}, and contribute to angular momentum transport during the post-dynamical phase \citep{Gagnier2024}. Accretion may therefore further strengthen the magnetic feedback on the system. These effects will be explored in detail in a forthcoming work \inprepcite{Gagnier}.
 Figure~\ref{fig:Drag_acc} shows the time evolution of the drag and radial forces acting on the companion, for a non-accreting simulation as well as accreting simulations with $\gamma = 100$ and $\delta = 0$ and $1$. We find that, while accretion induces strong temporal variability in both drag and radial forces, the resulting change in the mean drag force is likely not significant in view of the system's intrinsic stochastic fluctuations (see Appendix~\ref{app:stat}).  These force fluctuations translate directly into variations in the companion's instantaneous orbital decay rate, while the variable balance between drag and radial forces can excite eccentricity during the dynamical inspiral. The impact of both effects on the orbital evolution remains to be quantified. For clarity, the linear momentum accretion rate is omitted from the figure. It acts like a small ``negative'' drag force (or thrust), contributing only $\sim 0.27\%$ of the drag for $\delta = 0$ and $\sim 0.55\%$ for $\delta = 1$.

\subsection{Rotational effects}\label{sec:rot}
The Rossby number,
\begin{equation}
{\rm Ro} \sim \frac{|(\buu \cdot \bnabla) \buu|}{|\boldsymbol{\Omega} \times \buu|} \sim  
\begin{cases}
\displaystyle
\frac{r_2}{\widetilde{R_a}} = \frac{2q + (1+q)^2 \epsilon_\rho}{2q (1+q)^2 \epsilon_\rho} & \text{if } \epsilon_\rho > 0, \\[12pt]
\displaystyle
\frac{r_2}{R_a} & \text{otherwise},
\end{cases}
\label{eq:rossby_number}
\end{equation}
quantifies the dynamical importance of rotation in the flow. For typical common-envelope parameters, with a binary mass ratio $q = 0.1$ and a stratification parameter $\epsilon_\rho = \mathcal{O}(1)$, ${\rm Ro} = \mathcal{O}(1)$. This implies that rotational effects associated with the orbital motion cannot be neglected. Still, standard wind-tunnel simulations \citep[e.g.,][]{MacLeod2015, Thun2016, MM2017, De2020, Prust2024} ignore rotation entirely.

\begin{figure*}
    \centering
    \includegraphics[]{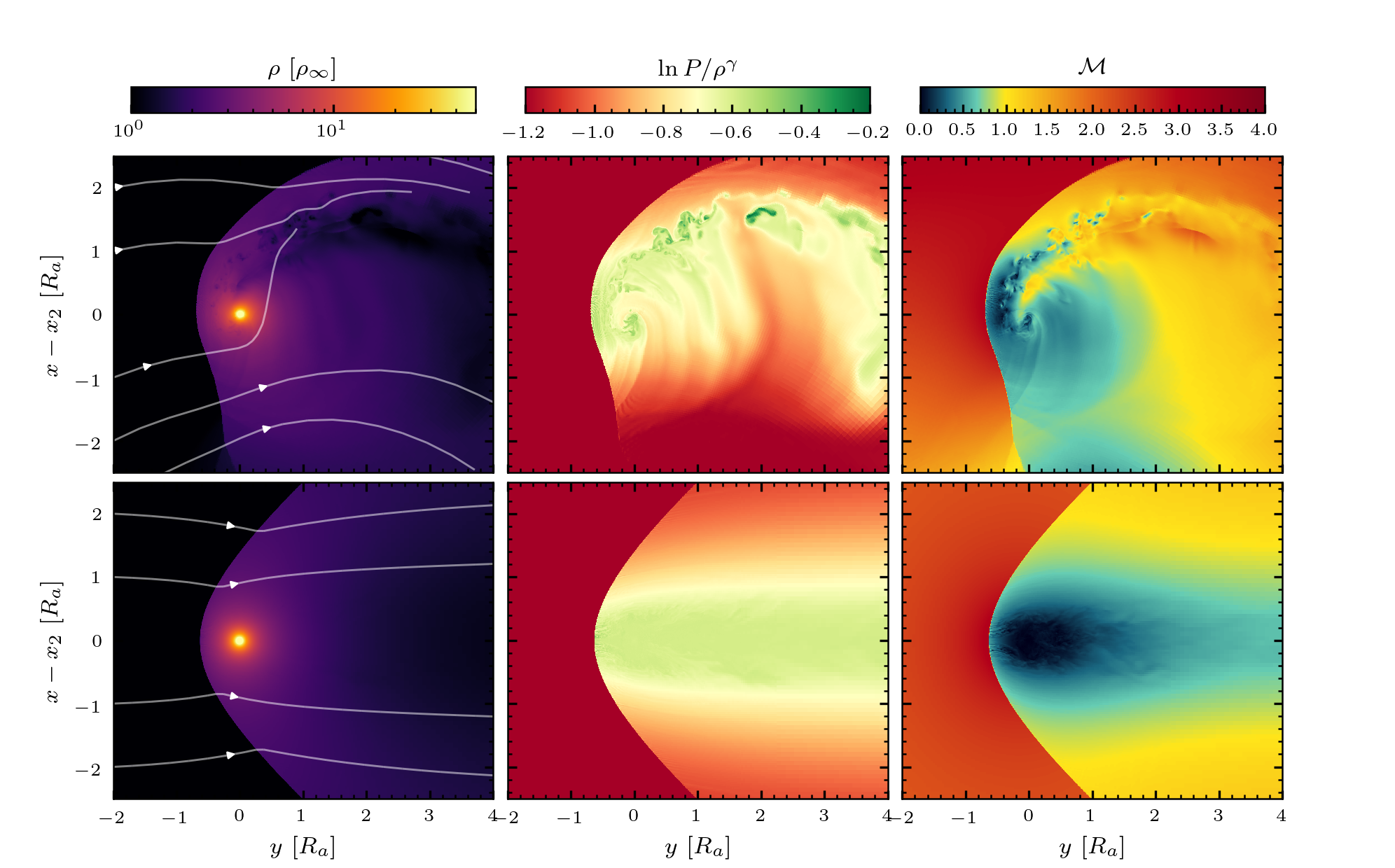}
\caption{Zoomed-in density, pseudo-entropy, and Mach number snapshots at $t = 100\, R_a/u_\infty$ in the $xy$ plane for non-stratified simulations with ${\rm Ro} = 5.5$ and $\mathcal{M}_\infty = 2$. The top panel includes rotation; the bottom does not. White lines indicate streamlines.}
\label{fig:rot}
\end{figure*}

\begin{figure}
    \centering
    \includegraphics[]{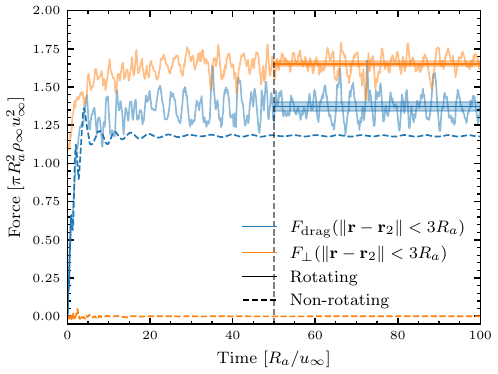}
    \caption{Time evolution of the radial and drag forces exerted by the gas on the companion in non-stratified simulations with ${\rm Ro} = 5.5$ and $\mathcal{M}_\infty = 2$, with and without rotation. Forces are integrated within a sphere of radius $3 R_a$. The shaded regions indicate the $3\sigma$ range (see Appendix \ref{app:stat}).}
    \label{fig:drag_rot}
\end{figure}

Rotation breaks the mirror symmetry of the flow, both dynamically via the Coriolis force and statically through the asymmetry of the effective gravito-centrifugal potential,
$\Phi_{\rm gc} = \Phi_1 + \Phi_2 - (\Omega \mathbf{e}_z \times \mathbf{r})^2/2$. 
In the orbiting frame with angular velocity $\Omega$, the azimuthal velocity of the inflowing gas increases linearly with distance from the rotation axis, further enhancing the symmetry breaking. As a result of this broken symmetry, inflowing material is deflected around the companion, generating $z$-directed angular momentum in the companion frame and producing a net positive lift force. This effect is illustrated in Fig.~\ref{fig:rot}, where we compare density snapshots from a Cartesian wind-tunnel and a rotating simulation on a spherical grid with ${\rm Ro} = 5.5$, at quasi-steady state. Both simulations are non-stratified, non-accreting, and have $\mathcal{M}_\infty = 2$ and $h_s = 0.1, R_a$. Figure~\ref{fig:drag_rot} shows the time evolution of the drag and lift forces. While rotation only slightly increases the drag force (by $\sim 14\%$ in this example), the lift force may significantly influence orbital contraction and excite eccentricity during the plunge-in phase (see also Sect.~\ref{sec:draglift}).
The flow also develops a secondary shock and strong, large-scale shear, which triggers shear instabilities that drive turbulent eddies. Numerical dissipation of these eddies generates entropy, heating the gas and driving the expansion of hot bubbles.

\subsection{Stratification effects}\label{sec:strat}
We performed hydrodynamical simulations with a fixed softening radius of $h_s = 0.05\, \widetilde{R_a}$. Unlike the previous sections, which focus on unstratified envelopes, we now explore the effects of stratification by varying the parameter $\epsilon_\rho$, thereby probing different depths within the primary's envelope (see Fig.\ref{fig:Prhoini}). In these stratified runs, the Mach number of the flow is directly determined by the value of $\epsilon_\rho$:
\begin{equation}
   \mathcal{M}_\infty^2 = \epsilon_\rho \frac{(1+q)^2}{2q} \ . 
\end{equation}  

\subsubsection{Flow morphology}
\begin{figure*}
    \centering
    \includegraphics[]{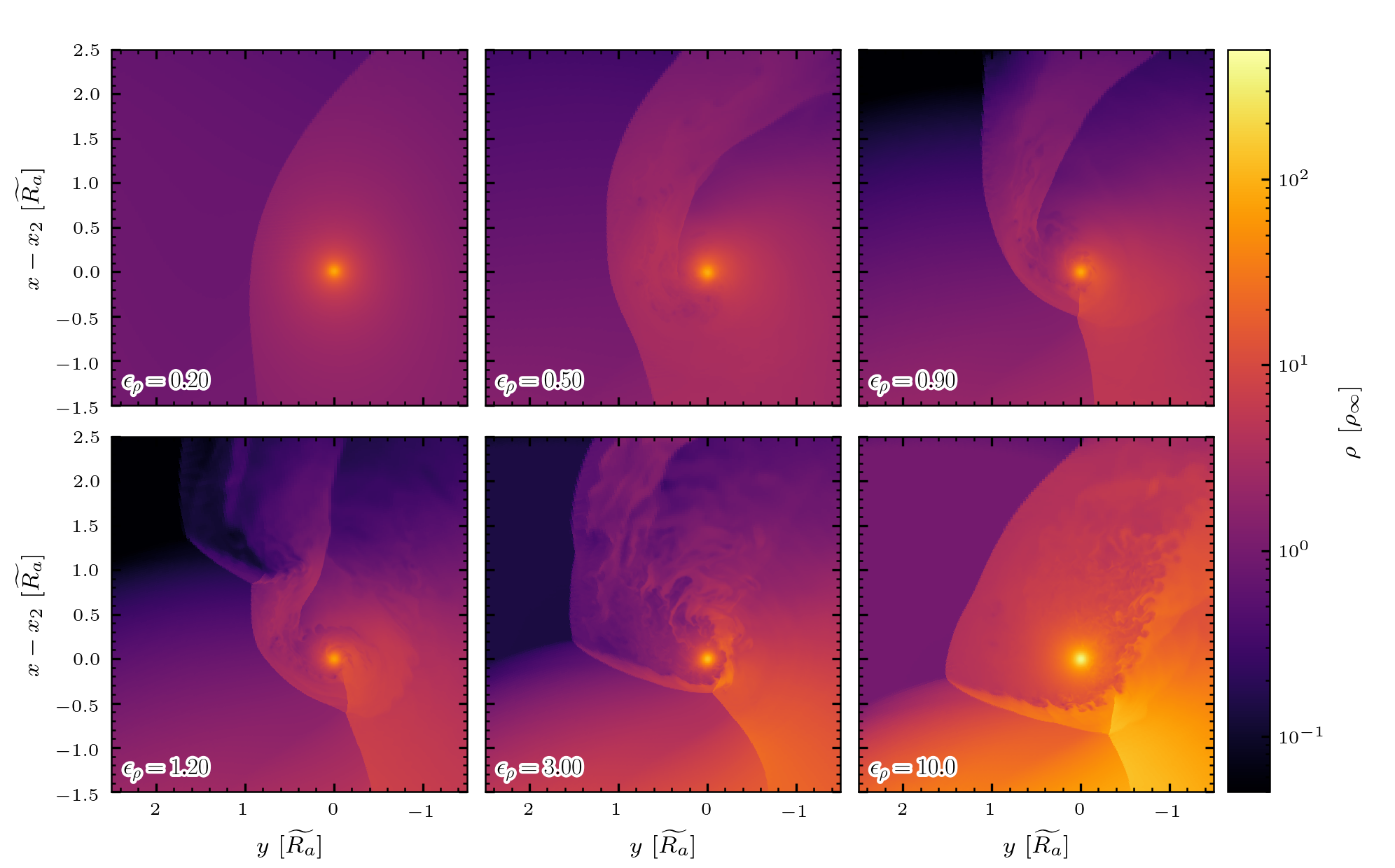}\\    \includegraphics[]{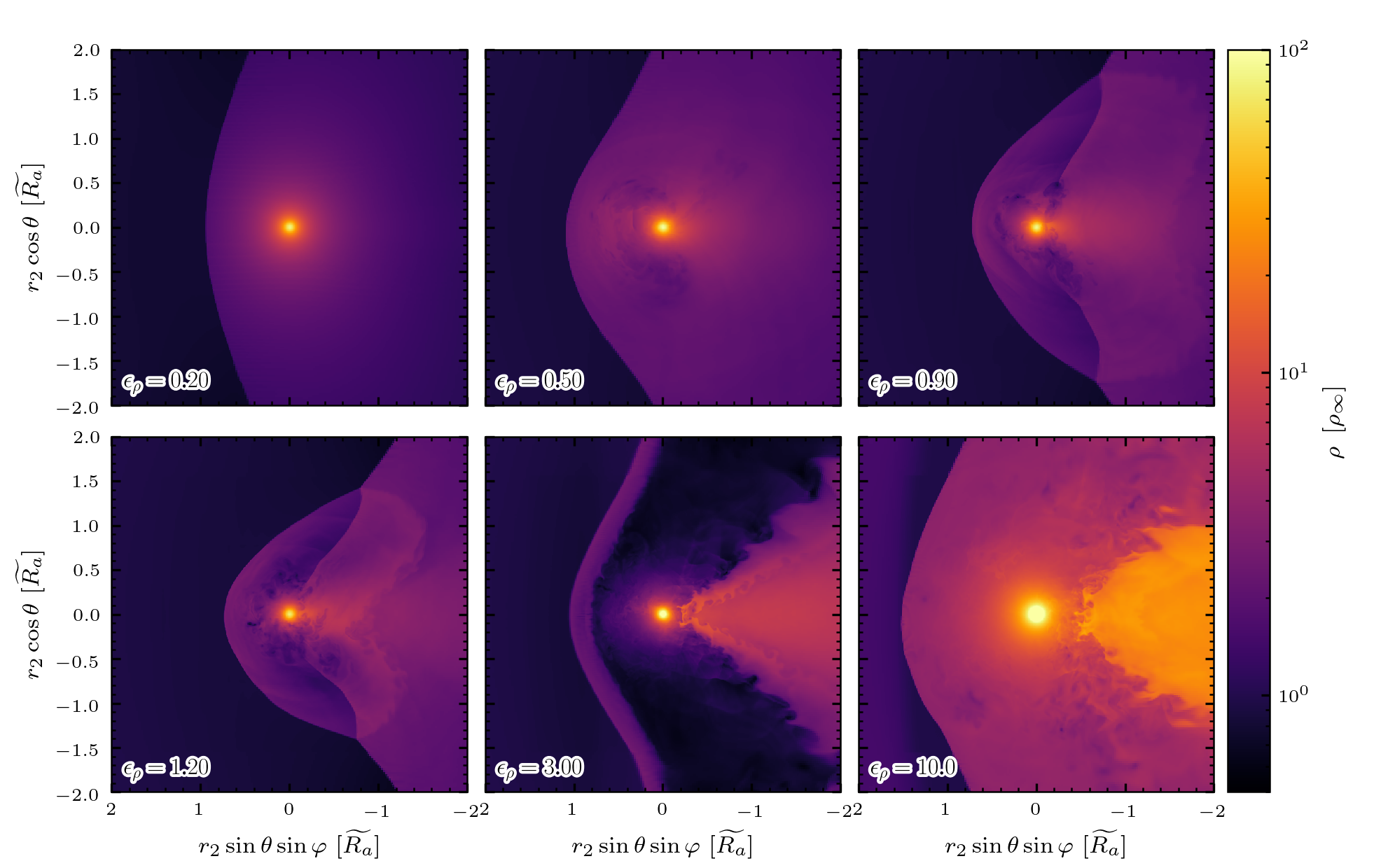}
\caption{
Zoomed-in snapshots of the density at $t = 100\ \widetilde{R_a}/u_\infty$, in the $xy$ orbital plane (first panel) and on a spherical shell at radius $r_2$ (second panel). }
    \label{fig:snapshots}
\end{figure*}

Figure~\ref{fig:snapshots} shows zoomed-in snapshots of density at $t = 100\,\widetilde{R_a}/u_\infty$, in the $xy$ orbital plane and projected on a spherical shell at radius $r_2$, for different values of the stratification parameter ($\epsilon_\rho$), including rotation but without accretion. Here, $\epsilon_\rho = 0.2$ corresponds to the innermost part of the $2\, M_\odot$ red giant envelope, while $\epsilon_\rho = 10$ corresponds to the outermost parts (see Fig.~\ref{fig:Prhoini}). In these simulations, the upstream flow is vertically stratified in density and pressure, and the local Mach number varies vertically due to the combined effects of the radial sound-speed gradient from stratification and the azimuthal velocity profile with constant angular velocity $\Omega$. The companion gravitationally deflects the surrounding gas, and the density stratification produces asymmetric gravitational focusing, with denser layers contributing more inertia, resulting in a vertically offset wake. Coriolis and centrifugal forces in the rotating frame further influence the vertical asymmetry of the wake (see Sect.~\ref{sec:rot}). The magnitude of the mirror asymmetry increases with the stratification parameter ($\epsilon_\rho$).
The post-shock radial stratification is shaped by the upstream density and pressure structure and the radial variation of the upstream velocity. This results in a corresponding radial variation in post-shock azimuthal momentum, which causes the gas to be differentially deflected by the companion and generates $z$-directed angular momentum in the companion's orbiting frame ($\be_z \cdot(\be_r^\ast \times \mathbf{u}^\ast) \neq 0$). For $\epsilon_\rho \gtrsim 3$, strong shear flows develop around the companion in the orbital plane and in the wake above and below it, leading to Kelvin-Helmholtz instabilities.

\subsubsection{Drag and lift forces}\label{sec:draglift}
\begin{figure}
    \centering
    \includegraphics[]{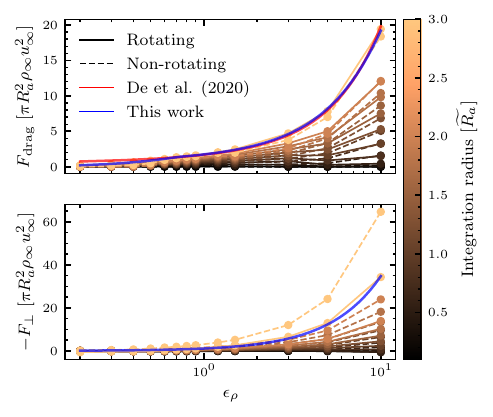}
\caption{
Time-averaged drag (top) and lift (bottom) forces on the companion as functions of the stratification parameter ($\epsilon_\rho$) and integration radius, averaged over $t \in [50,\,100]\,\widetilde{R_a} / u_\infty$. The red curve shows the prescription from \citet[Eq.~A4]{De2020}. The blue curve shows our fitting formulae (Eq.~\ref{eq:fit_fdrag}).}
\label{fig:drag_eps}
\end{figure}

Figure~\ref{fig:drag_eps} shows the time-averaged drag and radial forces exerted by the gas on the companion, averaged over $t \in [50,\,100]\,\widetilde{R_a} / u_\infty$ as functions of the stratification parameter ($\epsilon_\rho$) and of the radius of the integration volume. Similar to \cite{MM2017} and \cite{De2020}, we find that the drag force increases with increasing stratification strength. 
The fitting formula for the drag force proposed by \citet{De2020} provides an excellent match to our results in strongly stratified parts of the envelope ($\epsilon_\rho > 0.7$), but significantly overestimates the drag in more weakly stratified regions.
This difference arises because our simulations neglect accretion, so the measured drag reflects only the gravitational interaction with the stratified gas. In contrast, \citet{MM2017} and \citet{De2020} remove mass over a large region around the companion, transferring momentum from the gas to the object. While this mass removal generally reduces the net drag force \citep[e.g.,][and Sect.~\ref{sec:full}]{MM2017}, it also alters the large-scale density distribution, which can increase the total measured drag force. In strongly stratified flows, the accretion rate is lower (see Fig.~\ref{fig:MHL_prof} and Sect.~\ref{sec:full}), and the steep density gradients dominate the wake structure, reducing the discrepancy between the two approaches.
By neglecting accretion, we isolate the drag arising solely from the gravitational interaction with the stratified gas, without relying on sub-grid accretion prescriptions. Accretion effects are discussed separately in Sects.~\ref{sec:accr} and \ref{sec:full}. For $q = 0.1$, we provide an alternative fitting formula for the gravitational drag force in Eq.~\eqref{eq:fit_fdrag}.

Stratification also generates a lift force perpendicular to the companion's orbital motion. For steep density gradients, this force is directed inward, toward higher-density regions, and its magnitude increases with both $\epsilon_\rho$ and the size of the integration volume. However, for weak stratification ($\epsilon_\rho \lesssim 0.5$), the lift force reverses direction, producing a net outward contribution that remains too small to significantly affect the orbital evolution.
 As discussed in Sect.~\ref{sec:rot}, rotation has the opposite effect: it generates an outward force, directed toward lower-density regions. Including rotation therefore reduces the net inward force, especially in outer layers with high $\epsilon_\rho$ and thus low \text{Ro}. This is illustrated in the bottom panel of Fig.~\ref{fig:drag_eps}. For a binary mass ratio $ q = 0.1 $, we propose a simple fitting formula for this lift  force in Eq.~\eqref{eq:fit_fperp}.

\begin{subequations}\label{eq:fit}
\begin{align}
\log\frac{F_{\rm drag}(3\, \widetilde{R_a})}{\pi R_a^2 \rho_\infty u_\infty^2}
&\simeq 
\begin{cases}
\displaystyle\sum_{m=0}^2 a_m \mathcal{M}_\infty^{*m} + \left( \frac{ a_3}{\sqrt{\mathcal{M}_\infty}} -  \frac{ a_3}{\sqrt{\mathcal{M}_\infty^\ast}} \right),
& \displaystyle\frac{\mathcal{M}_\infty}{\mathcal{M}_\infty^{*}} \leq 1, \\[12pt]
\displaystyle\sum_{m=0}^2 a_m \, \mathcal{M}_\infty^m,
& \displaystyle\frac{\mathcal{M}_\infty}{\mathcal{M}_\infty^{*}} > 1,
\end{cases}
\label{eq:fit_fdrag}      
\end{align}
\begin{align}
\frac{F_{\perp}(3\, \widetilde{R_a})}{\pi R_a^2 \rho_\infty u_\infty^2}
&\simeq b_0 \mathcal{M}_\infty^3,
\label{eq:fit_fperp}
\end{align}
\end{subequations}
where $\mathcal{M}_\infty^{*} = \mathcal{M}_\infty(\epsilon_\rho=0.7)$. Fitting coefficients $a_m$ and $b_m$ are given in Table~\ref{tab:fit}. Convergence tests for $\epsilon_\rho = 3$ (Fig.~\ref{fig:Drag_rsoft}) show that both drag and lift forces converge for softening radii below roughly $\widetilde{R_a}$. The value used in this section, $h_s = 0.05,\widetilde{R_a}$, falls well within the converged regime. We expect this behavior to hold for other values of $\epsilon_\rho$.
We show the solution to a two-body problem that includes drag and lift forces acting on the secondary in Fig.~\ref{fig:orbit}. For that, we used the $\epsilon_\rho$ radial profile of the $2\,M_\odot$ red giant MESA profile (Fig.~\ref{fig:Prhoini}), and we assumed a mass ratio $q = 0.1$. The secondary star is subject to the gravity force from the enclosed primary mass and experiences additional drag and lift forces. We compared four models: (1) our full prescription with drag and lift forces (Eqs.~\ref{eq:fit_fdrag}--\ref{eq:fit_fperp}); (2) the same, but with $F_\perp = 0$; (3) the drag force prescription from \cite{De2020}; and (4) a standard Hoyle--Lyttleton  drag force $\pi R_a^2 \rho_\infty u_\infty^2$. We find that our model (Eqs.~\ref{eq:fit_fdrag}--\ref{eq:fit_fperp}) results in an initial inspiral that is faster than both the standard Hoyle--Lyttleton drag and the drag force prescription from \cite{De2020}. In the outer layers, where $\epsilon_\rho > 0.7$, the difference with \cite{De2020} arises from the inclusion of the (inward) lift force $F_\perp$. In the innermost layers, where stratification is weaker ($\epsilon_\rho < 0.7$), further differences stem from the stronger drag force in our model.

As in the non-stratified simulations, we find that the drag force follows the Coulomb-logarithm scaling \eqref{eq:Coulomb}, but with $b_{\rm min} \simeq 0.76\,\widetilde{R_a}$, and only for integration radii $b_{\rm max} \gtrsim 2\, \widetilde{R_a}$. 
Consequently, the fitting formula for the drag force \eqref{eq:fit_fdrag} can be rescaled for larger radii $b_{\rm max} > 3\,\widetilde{R_a}$ by
\begin{equation}
F_{\rm drag}(b_{\rm max}) \simeq F_{\rm drag}(3\,\widetilde{R_a}) \,
\frac{\ln (b_{\rm max}/(0.76\, \widetilde{R_a}))}{\ln (3.95)} \ ,
\end{equation}
keeping in mind that $b_{\rm max}$ is bound by the causal extent of the wake.
Conversely, the lift force does not follow a Coulomb-logarithm scaling, and its dependence on $b_{\rm max}$ varies with the stratification parameter, precluding a simple rescaling. Consequently, Eq.~\eqref{eq:fit_fperp} should be regarded as a lower bound for the actual lift force acting on the companion. This limitation complicates the ability of local simulations to accurately capture its role in orbital evolution and eccentricity excitation, and poses a challenge for devising reliable prescriptions for one-dimensional evolutionary codes.

\subsection{Combined effects of stratification, rotation, and accretion}\label{sec:full}
Here, we investigate the combined effects of stratification, rotation, and accretion. We consider three stratification parameters, $\epsilon_\rho = 0.3$, $0.5$, and $3$, which correspond approximately to $r_2/R_\ast \simeq 0.35$, $0.5$, and $0.9$, respectively. In this section, unless stated otherwise, we adopt $\gamma = 1000$ and $\delta = 1$, yielding mass accretion rates (in terms of $\dot{M}_{\rm HL}$) comparable to those of the non-stratified simulations with $\gamma = 100$ presented in Sect.~\ref{sec:accr}. The parameters and diagnostics of these simulations are summarized in Table~\ref{tab:rotstratacc}. 

\begin{figure}
    \centering
    \includegraphics[]{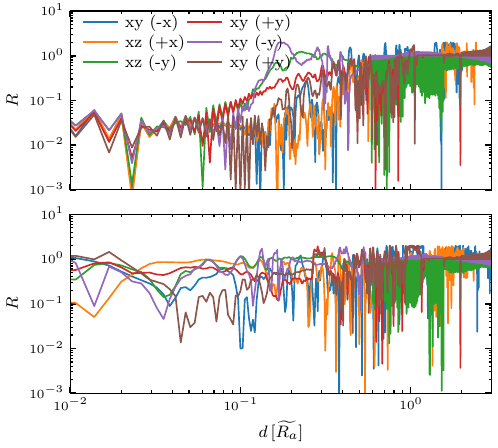}
\caption{Normalized hydrostatic-centrifugal balance residual
$R = \left\lVert \nabla P + \rho  \nabla \Phi_2 - (u_\varphi^\ast \sin \theta^\ast)^2/d\right\rVert/\max \left( \lVert \nabla P \rVert, \lVert \rho  \nabla \Phi_2 - (u_\varphi^\ast \sin \theta^\ast)^2/d \rVert \right)$
profiles along radial rays from the companion for rotating and stratified simulations with $\epsilon_\rho = 3$. Top panels: Non-accreting simulation. Bottom panels: Run with $\delta = 1$ and $\gamma = 1000$. $R$ is time-averaged within $t \in [99, 100]\, \widetilde{R_a} / u_\infty$.}
\label{fig:HEb}
\end{figure}

Figure~\ref{fig:HEb} shows the normalized hydrostatic-centrifugal balance residual profiles along radial rays from the companion for a non-accreting and an accreting simulation with $\epsilon_\rho = 3$, averaged over $t \in [99, 100]\, \widetilde{R_a} / u_\infty$. As also found in Sect.~\ref{sec:accr}, accretion prevents the formation of a quasi-hydrostatic bubble around the companion (see also Figs.~\ref{fig:snap_acc_strat_xy} and \ref{fig:snap_acc_strat_xz}). In the absence of accretion, such a bubble forms, with the centrifugal contribution to the force balance being less than $10\%$ of the gravitational force (not shown). Unlike the non-stratified, nonrotating case, the bubble does not extend to the shock front and instead has a radius of $\sim 0.1\, \widetilde{R_a}$. Quasi-hydrostatic equilibrium implies barotropicity and thus an absence of vorticity generation from baroclinic torque within the bubble. 

We verified that the radius of the quasi-hydrostatic bubble does not depend on the softening radius in the non-accreting simulation by repeating it with $h_s = 0.1$, $0.2$, and $0.5\, \widetilde{R_a}$. The bubble size remains unchanged for $h_s = 0.1\, \widetilde{R_a}$, but it does not form for $h \ge 0.2\, \widetilde{R_a}$. Increasing the softening radius weakens the hydrostatic support of the gas surrounding the companion, making it more susceptible to perturbations from the strong surrounding flow \citep{Gagnier2025}. This same strong rotational flow also limits the extent of the hydrostatic bubble to a fraction of the shock stand-off distance. By contrast, in the accreting case, the gas near the companion is baroclinic (see Fig.~\ref{fig:HE}), leading to vorticity generation that may enhance local magnetic field amplification, compared to non-accreting simulations. The same behavior is observed for both $\epsilon_\rho = 0.3$ and $0.5$. In these cases, however, the stratification-induced asymmetry is much smaller than for $\epsilon_\rho = 3$, resulting in weaker $z$-directed angular momentum in the companion frame. This results in reduced shear and turbulence, leading to a more stable flow in the companion's vicinity. In the absence of accretion, the quasi-hydrostatic bubble is therefore significantly larger for weaker envelope stratification. In particular, for both $\epsilon_\rho = 0.3$ and $0.5$, the bubble extends all the way to the shock front, similarly to the non-stratified and nonrotating simulations discussed in Sect.~\ref{sec:accr}. We did not investigate the dependence of the bubble size on the softening radius for the stratification parameters considered here.

Despite accretion's impact on turbulence, the drag and lift forces remain essentially unchanged (see  Table~\ref{tab:rotstratacc}). Still, accretion plays a crucial thermodynamic role. The inflow generates entropy through shocks and small-scale dissipation near the companion (e.g., Figs.~\ref{fig:snap_acc_strat_xy} and \ref{fig:snap_acc_strat_xz}), converting a fraction of the orbital energy into thermal energy. This heating is then advected downstream, contributing to the gravitational unbinding of the envelope.

We further find that the contribution of linear momentum accretion to the total drag force increases as the stratification parameter ($\epsilon_\rho$) decreases and appears to saturate in the weak-stratification regime (see Table~\ref{tab:rotstratacc}). Additional simulations are required to confirm this trend. The main reason for this increase toward lower $\epsilon_\rho$ is that regions with smaller stratification parameters correspond to denser parts of the envelope, where accretion rates are higher. A similar trend is observed for the radial force; however, it does not saturate in the asymptotic limit of weak stratification, since both the radial momentum accretion and the lift force should vanish as $\epsilon_\rho \rightarrow 0$. Indeed, in this limit, the Rossby number diverges and the combined effects of stratification and rotation on wake asymmetry disappear, yielding $\dot{p}_r \rightarrow 0$ and $F_\perp \rightarrow 0$. For finite $\epsilon_\rho$, linear momentum accretion remains finite, producing a net thrust and lift force.

We finally briefly investigated the effect of allowing angular momentum accretion onto the companion's spin by choosing $\delta = 1$. To this end, we assumed the companion to be a 0.2~$M_\odot$ white dwarf (WD) with a radius $R_2 \sim 0.02\,R_\odot$\footnote{The sink radius in our simulations is $r_s = 0.05\,\widetilde{R_a} \in [0.05,0.42]\,R_\odot$, i.e., much larger than the assumed WD radius.}. The specific angular momentum of the accreted material along the companion's $z$-axis reads
\begin{equation}
\ell_z = \frac{\dot{\mathbf{J}}_{\rm acc}^{\rm spin} \cdot \mathbf{e}_z^\ast}{\dot M} \ .
\end{equation}
We report this quantity normalized by the Keplerian value at the WD surface $\ell_{\rm K} = \sqrt{G M_2 R_2}$ in Table~\ref{tab:rotstratacc}. As expected from the increasing wake asymmetry with stronger stratification, $\ell_z$ increases with increasing $\epsilon_\rho$. For $\epsilon_\rho = 3$, $\ell_z$ is $\sim 80\%$ of the Keplerian value at the WD surface. Because $\ell_z$ is computed from the total angular momentum and mass accretion rates integrated over the sink region, it represents a weighted average over the accreted material. Hence, for $\epsilon_\rho = 3$, individual parcels of accreted gas can already carry specific angular momentum above the Keplerian value. At larger $\epsilon_\rho$, i.e., closer to the stellar surface, the global $\ell_z$ itself is expected to exceed the Keplerian limit. These results indicate that, even though the corresponding accretion rate is only about 1\% of the Hoyle--Lyttleton rate, the combination of parameters $\gamma = 1000$ and $\delta = 1$ likely overestimates the spin torque on the WD.
Assuming uniform rotation of the WD, as well as a radius of gyration $k = 0.2$ \citep{Andronov1990}, we can measure the corresponding spin-up timescale,
\begin{equation}
\tau_{\rm spin-up} = \frac{J_{\rm K}}{\dot{\mathbf{J}}_{\rm acc}^{\rm spin} \cdot \mathbf{e}_z^\ast}, \qquad
J_{\rm K} = k M_2 \sqrt{G M_2 R_2} \ .
\end{equation}
The values of $\tau_{\rm spin-up}$ are also reported in Table~\ref{tab:rotstratacc}. We find that for all three accreting simulations, even in the $\epsilon_\rho = 3$ case where part of the accreted material is likely rotating super-critically, the spin-up timescale of the WD remains much longer than the dynamical inspiral timescale, which is of order $1\,P_{\rm orb}$. This suggests that the WD's spin is unlikely to change significantly during the dynamical inspiral phase. Moreover, we find that the dependence of $\tau_{\rm spin-up}$ on $\epsilon_\rho$ is non-monotonic. This behavior arises because $\tau_{\rm spin-up} \propto (\ell_z \dot{M})^{-1}$, with $\ell_z$ increasing with $\epsilon_\rho$ while $\dot{M}$ decreases with increasing $\epsilon_\rho$. This suggests the existence of a critical stratification parameter $0.5 \lesssim \epsilon_\rho^{\rm c} \lesssim 3$ at which the spin-up timescale reaches a minimum for our specific simulations parameters. We finally note that magnetic torques could modify the spin-up timescale by transferring angular momentum between the WD and the surrounding material.

\section{Conclusion}\label{sec:conclusion}
We studied the effects of rotation, accretion, and stratification on the dynamical inspiral phase of CEE. To this end, we performed three-dimensional local simulations of a $0.2\,M_\odot$ compact companion plunging into the envelope of a $2\,M_\odot$ red giant in the companion's orbiting frame, using the \texttt{Athena++} code. Accretion onto the companion was modeled with either a standard sink, which allows angular momentum accretion onto the companion's spin, or a torque-free sink, which prevents it. The main results are as follows:

\begin{enumerate}
    \item The rotation associated with the orbital motion has a weak impact on the gravitational drag force. Measured drag forces are consistent with \cite{De2020} in the strong stratification regime. Asymmetries from the Coriolis force and the effective gravito-centrifugal potential generate a lift force, which partially counteracts the inward radial force from density stratification. The total radial force significantly affects the orbital evolution and cannot be neglected. We propose a new prescription for both drag and lift forces. Unlike the drag, the lift force does not follow a Coulomb-logarithm scaling, and the proposed lift force prescription should therefore be regarded as a lower limit.
    
    \item In the absence of accretion, a quasi-hydrostatic bubble forms around the companion. In the weak stratification regime, the bubble extends up to the shock front, while in the strong stratification regime, strong rotational flows in the companion's frame limit its radius. Such bubbles contribute negligibly to the net drag and lift forces and suppress vorticity growth. In the non-stratified and nonrotating regime, the bubble is never formally gravitationally bound and can be disrupted by small perturbations.
    
    \item Accretion prevents the formation of a quasi-hydrostatic bubble, producing weak shocks and strong turbulence near the companion. These processes generate entropy that is advected downstream and may aid envelope ejection and enhance magnetic energy amplification. Within the considered parameter space, accretion marginally affects drag and lift forces, remaining within the intrinsic stochastic variability of the flow. Using a standard sink, angular momentum accretion onto the companion's spin can be overestimated because the companion's physical size is much smaller than the sink radius. The companion's spin-up timescale remains much longer than the dynamical inspiral timescale. The spin-up rate varies non-monotonically with depth, peaking at intermediate radii where the product of accreted specific angular momentum, $\ell_z$, and accretion rate, $\dot{M}$, is largest.
\end{enumerate}

Magnetic fields influence CEE by affecting gas dynamics and orbital separation and contributing to the launching of bipolar outflows \citep[][]{Ohlmann2016b, Ondratschek2022, Gagnier2024, Vetter2024, Vetter2025}. Global magnetohydrodynamic simulations show that magnetic energy is predominantly amplified near the companion, but limited resolution near the companion restricts detailed study. Local simulations, such as those presented here, provide an ideal framework for examining magnetic amplification and its feedback on flow dynamics. Studies of massive objects in magnetized gas are sparse \citep[][]{Dokuchaev1964, Sanchez2012, Cunningham2012, Shadmehri2012, LeeBHL2014, kaaz2023}, and none have included the combined effects of rotation, stratification, and accretion. Addressing these effects is the focus of forthcoming work.

\begin{acknowledgements}
 D.G., G.L., M.V., R.A., and F.K.R. acknowledge support by the Klaus-
Tschira Foundation. D.G., M.V., R.A., and F.K.R. acknowledge funding by the European Union (ERC, ExCEED, project number 101096243). Views and opinions expressed are, however, those of the authors only and do not necessarily reflect those of the European Union or the European Research Council Executive Agency. Neither the European Union nor the granting authority can be held responsible for them. This work has received funding from the European Research Council (ERC) under the European Union's Horizon 2020 research and innovation program (Grant agreement No. 945806)   
\end{acknowledgements}

\bibliographystyle{aa}
\bibliography{bibnew}

@string{AN = "Astron. Nachr."}

@string{AA = "Astron. \& Astrophys."}

@string{AA = "A\&A"}

@string{ApJ = "ApJ"}

@string{ApJL = "ApJ Lett."}

@string{ApJS = "Astrophys. J. Supp. Ser."}

@string{M = "Mathematika"}

@string{MNRAS = "Mon. Not. R. astr. Soc."}

@string{MNRAS = "MNRAS"}

@string{N = "Nature"}

@string{PRD = "Phys. Rev. D"}

@preamble{ " \newcommand{\noop}[1]{} " }

@ARTICLE{kaaz2023,
       author = {{Kaaz}, Nicholas and {Schr{\o}der}, Sophie Lund and {Andrews}, Jeff J. and {Antoni}, Andrea and {Ramirez-Ruiz}, Enrico},
        title = "{The Hydrodynamic Evolution of Binary Black Holes Embedded within the Vertically Stratified Disks of Active Galactic Nuclei}",
      journal = {\apj},
         year = 2023,
        month = feb,
       volume = {944},
       number = {1},
          eid = {44},
        pages = {44},
          doi = {10.3847/1538-4357/aca967},
archivePrefix = {arXiv},
       eprint = {2103.12088},
 primaryClass = {astro-ph.HE},
       adsurl = {https://ui.adsabs.harvard.edu/abs/2023ApJ...944...44K}
}

@BOOK{Ivanova2020,
       author = {{Ivanova}, Natalia and {Justham}, Stephen and {Ricker}, Paul},
        title = "{Common Envelope Evolution}",
         year = 2020,
          doi = {10.1088/2514-3433/abb6f0},
       adsurl = {https://ui.adsabs.harvard.edu/abs/2020cee..book.....I}
}

@ARTICLE{Torres2025,
       author = {{Torres}, S. and {Gili}, M. and {Rebassa-Mansergas}, A. and {Santos-Garc{\'\i}a}, A. and {Brown}, A.~J. and {Parsons}, S.~G.},
        title = "{Reconstructing post-common envelope white dwarf{\textendash}main-sequence binary histories through inverse population synthesis techniques: A case study of ZTF eclipsing binaries}",
      journal = {\aap},
         year = 2025,
        month = jun,
       volume = {698},
          eid = {A173},
        pages = {A173},
          doi = {10.1051/0004-6361/202554039},
archivePrefix = {arXiv},
       eprint = {2505.01505},
 primaryClass = {astro-ph.SR},
       adsurl = {https://ui.adsabs.harvard.edu/abs/2025A&A...698A.173T}
}

@article{Ge2022,
doi = {10.3847/1538-4357/ac75d3},
url = {https://dx.doi.org/10.3847/1538-4357/ac75d3},
year = {2022},
month = {jul},
publisher = {The American Astronomical Society},
volume = {933},
number = {2},
pages = {137},
author = {Ge, Hongwei and Tout, Christopher A. and Chen, Xuefei and Kruckow, Matthias U. and Chen, Hailiang and Jiang, Dengkai and Li, Zhenwei and Liu, Zhengwei and Han, Zhanwen},
title = {The Common Envelope Evolution Outcome—A Case Study on Hot Subdwarf B Stars},
journal = {The Astrophysical Journal}
}

@ARTICLE{Antoni2019,
       author = {{Antoni}, Andrea and {MacLeod}, Morgan and {Ramirez-Ruiz}, Enrico},
        title = "{The Evolution of Binaries in a Gaseous Medium: Three-dimensional Simulations of Binary Bondi-Hoyle-Lyttleton Accretion}",
      journal = {\apj},
         year = 2019,
        month = oct,
       volume = {884},
       number = {1},
          eid = {22},
        pages = {22},
          doi = {10.3847/1538-4357/ab3466},
archivePrefix = {arXiv},
       eprint = {1901.07572},
 primaryClass = {astro-ph.HE},
       adsurl = {https://ui.adsabs.harvard.edu/abs/2019ApJ...884...22A}
}

@ARTICLE{Rosselli-Calderon2024,
       author = {{Rosselli-Calderon}, Alejandra and {Yarza}, Ricardo and {Murguia-Berthier}, Ariadna and {Rohoza}, Valeriia and {Everson}, Rosa Wallace and {Antoni}, Andrea and {MacLeod}, Morgan and {Ramirez-Ruiz}, Enrico},
        title = "{The Evolution of Binaries Embedded Within Common Envelopes}",
      journal = {\apj},
         year = 2024,
        month = dec,
       volume = {977},
       number = {1},
          eid = {16},
        pages = {16},
          doi = {10.3847/1538-4357/ad84ee},
archivePrefix = {arXiv},
       eprint = {2404.08037},
 primaryClass = {astro-ph.SR},
       adsurl = {https://ui.adsabs.harvard.edu/abs/2024ApJ...977...16R}
}

@ARTICLE{Monaghan1985,
       author = {{Monaghan}, J.~J. and {Lattanzio}, J.~C.},
        title = "{A refined particle method for astrophysical problems}",
      journal = {\aap},
         year = 1985,
        month = aug,
       volume = {149},
       number = {1},
        pages = {135-143},
       adsurl = {https://ui.adsabs.harvard.edu/abs/1985A&A...149..135M}
}

@ARTICLE{Dempsey2020,
       author = {{Dempsey}, Adam M. and {Mu{\~n}oz}, Diego and {Lithwick}, Yoram},
        title = "{Inner Boundary Condition in Quasi-Lagrangian Simulations of Accretion Disks}",
      journal = {\apjl},
         year = 2020,
        month = apr,
       volume = {892},
       number = {2},
          eid = {L29},
        pages = {L29},
          doi = {10.3847/2041-8213/ab800e},
archivePrefix = {arXiv},
       eprint = {2002.05164},
 primaryClass = {astro-ph.EP},
       adsurl = {https://ui.adsabs.harvard.edu/abs/2020ApJ...892L..29D}
}

@ARTICLE{Prust2024,
       author = {{Prust}, Logan J. and {Bildsten}, Lars},
        title = "{Flow morphology of a supersonic gravitating sphere}",
      journal = {\mnras},
         year = 2024,
        month = jan,
       volume = {527},
       number = {2},
        pages = {2869-2886},
          doi = {10.1093/mnras/stad3405},
archivePrefix = {arXiv},
       eprint = {2310.20315},
 primaryClass = {astro-ph.SR},
       adsurl = {https://ui.adsabs.harvard.edu/abs/2024MNRAS.527.2869P}
}

@ARTICLE{Scherbak2023,
       author = {{Scherbak}, Peter and {Fuller}, Jim},
        title = "{White dwarf binaries suggest a common envelope efficiency {\ensuremath{\alpha}}   1/3}",
      journal = {\mnras},
         year = 2023,
        month = jan,
       volume = {518},
       number = {3},
        pages = {3966-3984},
          doi = {10.1093/mnras/stac3313},
archivePrefix = {arXiv},
       eprint = {2211.02036},
 primaryClass = {astro-ph.SR},
       adsurl = {https://ui.adsabs.harvard.edu/abs/2023MNRAS.518.3966S}
}

@ARTICLE{DeMarco2011,
       author = {{De Marco}, Orsola and {Passy}, Jean-Claude and {Moe}, Maxwell and {Herwig}, Falk and {Mac Low}, Mordecai-Mark and {Paxton}, Bill},
        title = "{On the {\ensuremath{\alpha}} formalism for the common envelope interaction}",
      journal = {\mnras},
         year = 2011,
        month = mar,
       volume = {411},
       number = {4},
        pages = {2277-2292},
          doi = {10.1111/j.1365-2966.2010.17891.x10.48550/arXiv.1010.4374},
archivePrefix = {arXiv},
       eprint = {1010.4374},
 primaryClass = {astro-ph.SR},
       adsurl = {https://ui.adsabs.harvard.edu/abs/2011MNRAS.411.2277D}
}

@article{SHU1988,
title = {Efficient implementation of essentially non-oscillatory shock-capturing schemes},
journal = {Journal of Computational Physics},
volume = {77},
number = {2},
pages = {439-471},
year = {1988},
issn = {0021-9991},
doi = {https://doi.org/10.1016/0021-9991(88)90177-5},
url = {https://www.sciencedirect.com/science/article/pii/0021999188901775},
author = {Chi-Wang Shu and Stanley Osher}
}

@ARTICLE{Duffell2020,
       author = {{Duffell}, Paul C. and {D'Orazio}, Daniel and {Derdzinski}, Andrea and {Haiman}, Zoltan and {MacFadyen}, Andrew and {Rosen}, Anna L. and {Zrake}, Jonathan},
        title = "{Circumbinary Disks: Accretion and Torque as a Function of Mass Ratio and Disk Viscosity}",
      journal = {\apj},
         year = 2020,
        month = sep,
       volume = {901},
       number = {1},
          eid = {25},
        pages = {25},
          doi = {10.3847/1538-4357/abab95},
archivePrefix = {arXiv},
       eprint = {1911.05506},
 primaryClass = {astro-ph.SR},
       adsurl = {https://ui.adsabs.harvard.edu/abs/2020ApJ...901...25D}
}

@ARTICLE{Tiede2020,
       author = {{Tiede}, Christopher and {Zrake}, Jonathan and {MacFadyen}, Andrew and {Haiman}, Zoltan},
        title = "{Gas-driven Inspiral of Binaries in Thin Accretion Disks}",
      journal = {\apj},
         year = 2020,
        month = sep,
       volume = {900},
       number = {1},
          eid = {43},
        pages = {43},
          doi = {10.3847/1538-4357/aba432},
archivePrefix = {arXiv},
       eprint = {2005.09555},
 primaryClass = {astro-ph.GA},
       adsurl = {https://ui.adsabs.harvard.edu/abs/2020ApJ...900...43T}
}

@ARTICLE{Hirai2022,
       author = {{Hirai}, Ryosuke and {Mandel}, Ilya},
        title = "{A Two-stage Formalism for Common-envelope Phases of Massive Stars}",
      journal = {\apjl},
         year = 2022,
        month = oct,
       volume = {937},
       number = {2},
          eid = {L42},
        pages = {L42},
          doi = {10.3847/2041-8213/ac9519},
archivePrefix = {arXiv},
       eprint = {2209.05328},
 primaryClass = {astro-ph.SR},
       adsurl = {https://ui.adsabs.harvard.edu/abs/2022ApJ...937L..42H}
}

@ARTICLE{Nelemans2000,
       author = {{Nelemans}, G. and {Verbunt}, F. and {Yungelson}, L.~R. and {Portegies Zwart}, Simon F.},
        title = "{Reconstructing the evolution of double helium white dwarfs: envelope loss without spiral-in}",
      journal = {\aap},
         year = 2000,
        month = aug,
       volume = {360},
        pages = {1011-1018},
archivePrefix = {arXiv},
       eprint = {astro-ph/0006216},
 primaryClass = {astro-ph},
       adsurl = {https://ui.adsabs.harvard.edu/abs/2000A&A...360.1011N}
}

@ARTICLE{DiStefano2022,
       author = {{Di Stefano}, Rosanne and {Kruckow}, Matthias U. and {Gao}, Yan and {Neunteufel}, Patrick G. and {Kobayashi}, Chiaki},
        title = "{SCATTER: A New Common Envelope Formalism}",
      journal = {\apj},
         year = 2023,
        month = feb,
       volume = {944},
       number = {1},
          eid = {87},
        pages = {87},
          doi = {10.3847/1538-4357/acae9b},
archivePrefix = {arXiv},
       eprint = {2212.06770},
 primaryClass = {astro-ph.HE},
       adsurl = {https://ui.adsabs.harvard.edu/abs/2023ApJ...944...87D}
}

@ARTICLE{Webbink1984,
       author = {{Webbink}, R.~F.},
        title = "{Double white dwarfs as progenitors of R Coronae Borealis stars and type I supernovae.}",
      journal = {\apj},
         year = 1984,
        month = feb,
       volume = {277},
        pages = {355-360},
          doi = {10.1086/161701},
       adsurl = {https://ui.adsabs.harvard.edu/abs/1984ApJ...277..355W}
}

@ARTICLE{Trani2022,
       author = {{Trani}, Alessandro Alberto and {Rieder}, Steven and {Tanikawa}, Ataru and {Iorio}, Giuliano and {Martini}, Riccardo and {Karelin}, Georgii and {Glanz}, Hila and {Portegies Zwart}, Simon},
        title = "{Revisiting the common envelope evolution in binary stars: A new semianalytic model for N -body and population synthesis codes}",
      journal = {\prd},
         year = 2022,
        month = aug,
       volume = {106},
       number = {4},
          eid = {043014},
        pages = {043014},
          doi = {10.1103/PhysRevD.106.043014},
archivePrefix = {arXiv},
       eprint = {2205.13537},
 primaryClass = {astro-ph.SR},
       adsurl = {https://ui.adsabs.harvard.edu/abs/2022PhRvD.106d3014T}
}

@ARTICLE{Lau2025,
       author = {{Lau}, Mike Y.~M. and {Hirai}, Ryosuke and {Price}, Daniel J. and {Mandel}, Ilya and {Bate}, Matthew R.},
        title = "{Common envelopes in massive stars: III. The obstructive role of radiation transport in envelope ejection}",
      journal = {\aap},
         year = 2025,
        month = jul,
       volume = {699},
          eid = {A274},
        pages = {A274},
          doi = {10.1051/0004-6361/202554782},
archivePrefix = {arXiv},
       eprint = {2503.20506},
 primaryClass = {astro-ph.SR},
       adsurl = {https://ui.adsabs.harvard.edu/abs/2025A&A...699A.274L}
}

@ARTICLE{Roepke2023,
       author = {{R{\"o}pke}, Friedrich K. and {De Marco}, Orsola},
        title = "{Simulations of common-envelope evolution in binary stellar systems: physical models and numerical techniques}",
      journal = {Living Reviews in Computational Astrophysics},
         year = 2023,
        month = dec,
       volume = {9},
       number = {1},
          eid = {2},
        pages = {2},
          doi = {10.1007/s41115-023-00017-x},
archivePrefix = {arXiv},
       eprint = {2212.07308},
 primaryClass = {astro-ph.SR},
       adsurl = {https://ui.adsabs.harvard.edu/abs/2023LRCA....9....2R}
}

@ARTICLE{Bhattacharyya2025,
       author = {{Bhattacharyya}, Soumik and {Chamandy}, Luke and {Blackman}, Eric G. and {Frank}, Adam and {Liu}, Baowei},
        title = "{Understanding the Drag Torque in Common Envelope Evolution}",
      journal = {arXiv e-prints},
         year = 2025,
        month = jun,
          eid = {arXiv:2506.19547},
        pages = {arXiv:2506.19547},
          doi = {10.48550/arXiv.2506.19547},
archivePrefix = {arXiv},
       eprint = {2506.19547},
 primaryClass = {astro-ph.SR},
       adsurl = {https://ui.adsabs.harvard.edu/abs/2025arXiv250619547B}
}

@ARTICLE{Livio1988,
       author = {{Livio}, Mario and {Soker}, Noam},
        title = "{The Common Envelope Phase in the Evolution of Binary Stars}",
      journal = APJ,
         year = 1988,
        month = jun,
       volume = {329},
        pages = {764},
          doi = {10.1086/166419},
       adsurl = {https://ui.adsabs.harvard.edu/abs/1988ApJ...329..764L}
}

@ARTICLE{Dittmann2021,
       author = {{Dittmann}, Alexander J. and {Ryan}, Geoffrey},
        title = "{Preventing Anomalous Torques in Circumbinary Accretion Simulations}",
      journal = APJ,
         year = 2021,
        month = nov,
       volume = {921},
       number = {1},
          eid = {71},
        pages = {71},
          doi = {10.3847/1538-4357/ac1bbd},
archivePrefix = {arXiv},
       eprint = {2102.05684},
 primaryClass = {astro-ph.HE},
       adsurl = {https://ui.adsabs.harvard.edu/abs/2021ApJ...921...71D}
}

@article{Ablimit2016,
	doi = {10.3847/0004-637x/826/1/53},
	url = {https://doi.org/10.3847/0004-637x/826/1/53},
	year = 2016,
	month = {jul},
	publisher = {American Astronomical Society},
	volume = {826},
	number = {1},
	pages = {53},
	author = {Iminhaji Ablimit and Keiichi Maeda and Xiang-Dong Li},
	title = {{MONTE} {CARLO} {POPULATION} {SYNTHESIS} {OF} {POST}-{COMMON}-{ENVELOPE} {WHITE} {DWARF} {BINARIES} {AND} {TYPE} Ia {SUPERNOVA} {RATE}},
	journal = APJ
}

@ARTICLE{Munoz2019,
       author = {{Mu{\~n}oz}, Diego J. and {Miranda}, Ryan and {Lai}, Dong},
        title = "{Hydrodynamics of Circumbinary Accretion: Angular Momentum Transfer and Binary Orbital Evolution}",
      journal = APJ,
         year = 2019,
        month = jan,
       volume = {871},
       number = {1},
          eid = {84},
        pages = {84},
          doi = {10.3847/1538-4357/aaf867},
archivePrefix = {arXiv},
       eprint = {1810.04676},
 primaryClass = {astro-ph.HE},
       adsurl = {https://ui.adsabs.harvard.edu/abs/2019ApJ...871...84M}
}

@ARTICLE{Everson2020,
       author = {{Everson}, Rosa Wallace and {MacLeod}, Morgan and {De}, Soumi and {Macias}, Phillip and {Ramirez-Ruiz}, Enrico},
        title = "{Common Envelope Wind Tunnel: Range of Applicability and Self-similarity in Realistic Stellar Envelopes}",
      journal = {\apj},
         year = 2020,
        month = aug,
       volume = {899},
       number = {1},
          eid = {77},
        pages = {77},
          doi = {10.3847/1538-4357/aba75c},
archivePrefix = {arXiv},
       eprint = {2006.07471},
 primaryClass = {astro-ph.SR},
       adsurl = {https://ui.adsabs.harvard.edu/abs/2020ApJ...899...77E}
}

@ARTICLE{Ohlmann2016b,
       author = {{Ohlmann}, Sebastian T. and {R{\"o}pke}, Friedrich K. and {Pakmor}, R{\"u}diger and {Springel}, Volker and {M{\"u}ller}, Ewald},
        title = "{Magnetic field amplification during the common envelope phase}",
      journal = {\mnras},
         year = 2016,
        month = oct,
       volume = {462},
       number = {1},
        pages = {L121-L125},
          doi = {10.1093/mnrasl/slw144},
archivePrefix = {arXiv},
       eprint = {1607.05996},
 primaryClass = {astro-ph.SR},
       adsurl = {https://ui.adsabs.harvard.edu/abs/2016MNRAS.462L.121O}
}

@ARTICLE{HL1939,
       author = {{Hoyle}, F. and {Lyttleton}, R.~A.},
        title = "{The effect of interstellar matter on climatic variation}",
      journal = {Proceedings of the Cambridge Philosophical Society},
         year = 1939,
        month = jan,
       volume = {35},
       number = {3},
        pages = {405},
          doi = {10.1017/S0305004100021150},
       adsurl = {https://ui.adsabs.harvard.edu/abs/1939PCPS...35..405H}
}

@ARTICLE{LeeBHL2014,
       author = {{Lee}, Aaron T. and {Cunningham}, Andrew J. and {McKee}, Christopher F. and {Klein}, Richard I.},
        title = "{Bondi-Hoyle Accretion in an Isothermal Magnetized Plasma}",
      journal = {\apj},
         year = 2014,
        month = mar,
       volume = {783},
       number = {1},
          eid = {50},
        pages = {50},
          doi = {10.1088/0004-637X/783/1/50},
archivePrefix = {arXiv},
       eprint = {1401.7010},
 primaryClass = {astro-ph.GA},
       adsurl = {https://ui.adsabs.harvard.edu/abs/2014ApJ...783...50L}
}

@ARTICLE{Cunningham2012,
       author = {{Cunningham}, Andrew J. and {McKee}, Christopher F. and {Klein}, Richard I. and {Krumholz}, Mark R. and {Teyssier}, Romain},
        title = "{Radiatively Efficient Magnetized Bondi Accretion}",
      journal = {\apj},
         year = 2012,
        month = jan,
       volume = {744},
       number = {2},
          eid = {185},
        pages = {185},
          doi = {10.1088/0004-637X/744/2/185},
archivePrefix = {arXiv},
       eprint = {1201.0816},
 primaryClass = {astro-ph.SR},
       adsurl = {https://ui.adsabs.harvard.edu/abs/2012ApJ...744..185C}
}

@ARTICLE{PPM08,
       author = {{Colella}, Phillip and {Sekora}, Michael D.},
        title = "{A limiter for PPM that preserves accuracy at smooth extrema}",
      journal = {Journal of Computational Physics},
         year = 2008,
        month = jul,
       volume = {227},
       number = {15},
        pages = {7069-7076},
          doi = {10.1016/j.jcp.2008.03.034},
       adsurl = {https://ui.adsabs.harvard.edu/abs/2008JCoPh.227.7069C}
}

@ARTICLE{lhll,
       author = {{Minoshima}, Takashi and {Miyoshi}, Takahiro},
        title = "{A low-dissipation HLLD approximate Riemann solver for a very wide range of Mach numbers}",
      journal = {Journal of Computational Physics},
         year = 2021,
        month = aug,
       volume = {446},
          eid = {110639},
        pages = {110639},
          doi = {10.1016/j.jcp.2021.110639},
archivePrefix = {arXiv},
       eprint = {2108.04991},
 primaryClass = {physics.comp-ph},
       adsurl = {https://ui.adsabs.harvard.edu/abs/2021JCoPh.44610639M}
}

@ARTICLE{PLUTO2012,
       author = {{Mignone}, A. and {Zanni}, C. and {Tzeferacos}, P. and {van Straalen}, B. and {Colella}, P. and {Bodo}, G.},
        title = "{The PLUTO Code for Adaptive Mesh Computations in Astrophysical Fluid Dynamics}",
      journal = {\apjs},
         year = 2012,
        month = jan,
       volume = {198},
       number = {1},
          eid = {7},
        pages = {7},
          doi = {10.1088/0067-0049/198/1/7},
archivePrefix = {arXiv},
       eprint = {1110.0740},
 primaryClass = {astro-ph.HE},
       adsurl = {https://ui.adsabs.harvard.edu/abs/2012ApJS..198....7M}
}

@ARTICLE{Lohner1987,
       author = {{Lohner}, Rainald and {Morgan}, Ken and {Peraire}, Jaime and {Vahdati}, Mehdi},
        title = "{Finite element flux-corrected transport (FEM-FCT) for the Euler and Navier-Stokes equations}",
      journal = {International Journal for Numerical Methods in Fluids},
         year = 1987,
        month = oct,
       volume = {7},
        pages = {1093-1109},
          doi = {10.1002/fld.1650071007},
       adsurl = {https://ui.adsabs.harvard.edu/abs/1987IJNMF...7.1093L}
}

@ARTICLE{Chamandy2018,
       author = {{Chamandy}, Luke and {Frank}, Adam and {Blackman}, Eric G. and {Carroll-Nellenback}, Jonathan and {Liu}, Baowei and {Tu}, Yisheng and {Nordhaus}, Jason and {Chen}, Zhuo and {Peng}, Bo},
        title = "{Accretion in common envelope evolution}",
      journal = MNRAS,
         year = 2018,
        month = oct,
       volume = {480},
       number = {2},
        pages = {1898-1911},
          doi = {10.1093/mnras/sty1950},
archivePrefix = {arXiv},
       eprint = {1805.03607},
 primaryClass = {astro-ph.SR},
       adsurl = {https://ui.adsabs.harvard.edu/abs/2018MNRAS.480.1898C}
}

@ARTICLE{Hernquist1989,
       author = {{Hernquist}, Lars and {Katz}, Neal},
        title = "{TREESPH: A Unification of SPH with the Hierarchical Tree Method}",
      journal = {\apjs},
         year = 1989,
        month = jun,
       volume = {70},
        pages = {419},
          doi = {10.1086/191344},
       adsurl = {https://ui.adsabs.harvard.edu/abs/1989ApJS...70..419H}
}

@ARTICLE{Ricker2012,
       author = {{Ricker}, Paul M. and {Taam}, Ronald E.},
        title = "{An AMR Study of the Common-envelope Phase of Binary Evolution}",
      journal = APJ,
         year = 2012,
        month = feb,
       volume = {746},
       number = {1},
          eid = {74},
        pages = {74},
          doi = {10.1088/0004-637X/746/1/74},
archivePrefix = {arXiv},
       eprint = {1107.3889},
 primaryClass = {astro-ph.SR},
       adsurl = {https://ui.adsabs.harvard.edu/abs/2012ApJ...746...74R}
}

@ARTICLE{De2020,
       author = {{De}, Soumi and {MacLeod}, Morgan and {Everson}, Rosa Wallace and {Antoni}, Andrea and {Mandel}, Ilya and {Ramirez-Ruiz}, Enrico},
        title = "{Common Envelope Wind Tunnel: The Effects of Binary Mass Ratio and Implications for the Accretion-driven Growth of LIGO Binary Black Holes}",
      journal = APJ,
         year = 2020,
        month = jul,
       volume = {897},
       number = {2},
          eid = {130},
        pages = {130},
          doi = {10.3847/1538-4357/ab9ac6},
archivePrefix = {arXiv},
       eprint = {1910.13333},
 primaryClass = {astro-ph.SR},
       adsurl = {https://ui.adsabs.harvard.edu/abs/2020ApJ...897..130D}
}

@ARTICLE{MacLeod2015,
       author = {{MacLeod}, Morgan and {Ramirez-Ruiz}, Enrico},
        title = "{Asymmetric Accretion Flows within a Common Envelope}",
      journal = APJ,
         year = 2015,
        month = apr,
       volume = {803},
       number = {1},
          eid = {41},
        pages = {41},
          doi = {10.1088/0004-637X/803/1/41},
archivePrefix = {arXiv},
       eprint = {1410.3823},
 primaryClass = {astro-ph.SR},
       adsurl = {https://ui.adsabs.harvard.edu/abs/2015ApJ...803...41M}
}

@ARTICLE{Ostriker1999,
       author = {{Ostriker}, Eve C.},
        title = "{Dynamical Friction in a Gaseous Medium}",
      journal = APJ,
         year = 1999,
        month = mar,
       volume = {513},
       number = {1},
        pages = {252-258},
          doi = {10.1086/306858},
archivePrefix = {arXiv},
       eprint = {astro-ph/9810324},
 primaryClass = {astro-ph},
       adsurl = {https://ui.adsabs.harvard.edu/abs/1999ApJ...513..252O}
}

@ARTICLE{MacLeod2018,
       author = {{MacLeod}, Morgan and {Ostriker}, Eve C. and {Stone}, James M.},
        title = "{Runaway Coalescence at the Onset of Common Envelope Episodes}",
      journal = APJ,
         year = 2018,
        month = aug,
       volume = {863},
       number = {1},
          eid = {5},
        pages = {5},
          doi = {10.3847/1538-4357/aacf08},
archivePrefix = {arXiv},
       eprint = {1803.03261},
 primaryClass = {astro-ph.SR},
       adsurl = {https://ui.adsabs.harvard.edu/abs/2018ApJ...863....5M}
}

@ARTICLE{Stone2020,
       author = {{Stone}, James M. and {Tomida}, Kengo and {White}, Christopher J. and {Felker}, Kyle G.},
        title = "{The Athena++ Adaptive Mesh Refinement Framework: Design and Magnetohydrodynamic Solvers}",
      journal = {\apjs},
         year = 2020,
        month = jul,
       volume = {249},
       number = {1},
          eid = {4},
        pages = {4},
          doi = {10.3847/1538-4365/ab929b},
archivePrefix = {arXiv},
       eprint = {2005.06651},
 primaryClass = {astro-ph.IM},
       adsurl = {https://ui.adsabs.harvard.edu/abs/2020ApJS..249....4S}
}

@ARTICLE{Iben1984,
       author = {{Iben}, I., Jr. and {Tutukov}, A.~V.},
        title = "{Supernovae of type I as end products of the evolution of binaries with components of moderate initial mass.}",
      journal = {\apjs},
         year = 1984,
        month = feb,
       volume = {54},
        pages = {335-372},
          doi = {10.1086/190932},
       adsurl = {https://ui.adsabs.harvard.edu/abs/1984ApJS...54..335I}
}

@INPROCEEDINGS{Tutukov1979,
       author = {{Tutukov}, A. and {Yungelson}, L.},
        title = "{Evolution of massive common envelope binaries and mass loss.}",
    booktitle = {Mass Loss and Evolution of O-Type Stars},
         year = 1979,
       editor = {{Conti}, P.~S. and {De Loore}, C.~W.~H.},
       series = {IAU Symposium},
       volume = {83},
        month = jan,
        pages = {401-406},
       adsurl = {https://ui.adsabs.harvard.edu/abs/1979IAUS...83..401T}
}

@ARTICLE{Belczynski2005,
       author = {{Belczynski}, Krzysztof and {Bulik}, Tomasz and {Ruiter}, Ashley J.},
        title = "{New Constraints on Type Ia Supernova Progenitor Models}",
      journal = APJ,
         year = 2005,
        month = aug,
       volume = {629},
       number = {2},
        pages = {915-921},
          doi = {10.1086/431578},
archivePrefix = {arXiv},
       eprint = {astro-ph/0502196},
 primaryClass = {astro-ph},
       adsurl = {https://ui.adsabs.harvard.edu/abs/2005ApJ...629..915B}
}

@ARTICLE{Quirk1994,
       author = {{Quirk}, James J.},
        title = "{A contribution to the great Riemann solver debate}",
      journal = {International Journal for Numerical Methods in Fluids},
         year = 1994,
        month = mar,
       volume = {18},
       number = {6},
        pages = {555-574},
          doi = {10.1002/fld.1650180603},
       adsurl = {https://ui.adsabs.harvard.edu/abs/1994IJNMF..18..555Q}
}

@ARTICLE{Ondratschek2022,
       author = {{Ondratschek}, Patrick A. and {R{\"o}pke}, Friedrich K. and {Schneider}, Fabian R.~N. and {Fendt}, Christian and {Sand}, Christian and {Ohlmann}, Sebastian T. and {Pakmor}, R{\"u}diger and {Springel}, Volker},
        title = "{Bipolar planetary nebulae from common-envelope evolution of binary stars}",
      journal = {\aap},
         year = 2022,
        month = apr,
       volume = {660},
          eid = {L8},
        pages = {L8},
          doi = {10.1051/0004-6361/202142478},
archivePrefix = {arXiv},
       eprint = {2110.13177},
 primaryClass = {astro-ph.SR},
       adsurl = {https://ui.adsabs.harvard.edu/abs/2022A&A...660L...8O}
}

@ARTICLE{Gagnier2023,
       author = {{Gagnier}, Damien and {Pejcha}, Ond{\v{r}}ej},
        title = "{Post-dynamical inspiral phase of common envelope evolution. Binary orbit evolution and angular momentum transport}",
      journal = {\aap},
         year = 2023,
        month = jun,
       volume = {674},
          eid = {A121},
        pages = {A121},
          doi = {10.1051/0004-6361/202346057},
archivePrefix = {arXiv},
       eprint = {2302.00691},
 primaryClass = {astro-ph.SR},
       adsurl = {https://ui.adsabs.harvard.edu/abs/2023A&A...674A.121G}
}

@ARTICLE{Gagnier2024,
       author = {{Gagnier}, Damien and {Pejcha}, Ond{\v{r}}ej},
        title = "{Post-dynamical inspiral phase of common envelope evolution. The role of magnetic fields}",
      journal = {\aap},
         year = 2024,
        month = mar,
       volume = {683},
          eid = {A4},
        pages = {A4},
          doi = {10.1051/0004-6361/202348383},
archivePrefix = {arXiv},
       eprint = {2310.16880},
 primaryClass = {astro-ph.SR},
       adsurl = {https://ui.adsabs.harvard.edu/abs/2024A&A...683A...4G}
}

@ARTICLE{Gagnier2025,
       author = {{Gagnier}, Damien and {Pejcha}, Ond{\v{r}}ej},
        title = "{Journey to the center of the common envelope evolution. Inner dynamics of the post-dynamical inspiral}",
      journal = {\aap},
         year = 2025,
        month = apr,
       volume = {697},
          eid = {A68},
        pages = {A68},
          doi = {10.1051/0004-6361/202452616},
archivePrefix = {arXiv},
       eprint = {2412.04419},
 primaryClass = {astro-ph.SR},
       adsurl = {https://ui.adsabs.harvard.edu/abs/2024arXiv241204419G}
}

@ARTICLE{Andrassy2022,
       author = {{Andrassy}, R. and {Higl}, J. and {Mao}, H. and {Moc{\'a}k}, M. and {Vlaykov}, D.~G. and {Arnett}, W.~D. and {Baraffe}, I. and {Campbell}, S.~W. and {Constantino}, T. and {Edelmann}, P.~V.~F. and {Goffrey}, T. and {Guillet}, T. and {Herwig}, F. and {Hirschi}, R. and {Horst}, L. and {Leidi}, G. and {Meakin}, C. and {Pratt}, J. and {Rizzuti}, F. and {R{\"o}pke}, F.~K. and {Woodward}, P.},
        title = "{Dynamics in a stellar convective layer and at its boundary: Comparison of five 3D hydrodynamics codes}",
      journal = {\aap},
         year = 2022,
        month = mar,
       volume = {659},
          eid = {A193},
        pages = {A193},
          doi = {10.1051/0004-6361/202142557},
archivePrefix = {arXiv},
       eprint = {2111.01165},
 primaryClass = {astro-ph.SR},
       adsurl = {https://ui.adsabs.harvard.edu/abs/2022A&A...659A.193A}
}

@ARTICLE{Fragos2019,
       author = {{Fragos}, Tassos and {Andrews}, Jeff J. and {Ramirez-Ruiz}, Enrico and {Meynet}, Georges and {Kalogera}, Vicky and {Taam}, Ronald E. and {Zezas}, Andreas},
        title = "{The Complete Evolution of a Neutron-star Binary through a Common Envelope Phase Using 1D Hydrodynamic Simulations}",
      journal = APJL,
         year = 2019,
        month = oct,
       volume = {883},
       number = {2},
          eid = {L45},
        pages = {L45},
          doi = {10.3847/2041-8213/ab40d1},
archivePrefix = {arXiv},
       eprint = {1907.12573},
 primaryClass = {astro-ph.HE},
       adsurl = {https://ui.adsabs.harvard.edu/abs/2019ApJ...883L..45F}
}

@ARTICLE{Zorotovic2010,
       author = {{Zorotovic}, M. and {Schreiber}, M.~R. and {G{\"a}nsicke}, B.~T. and {Nebot G{\'o}mez-Mor{\'a}n}, A.},
        title = "{Post-common-envelope binaries from SDSS. IX: Constraining the common-envelope efficiency}",
      journal = {\aap},
         year = 2010,
        month = sep,
       volume = {520},
          eid = {A86},
        pages = {A86},
          doi = {10.1051/0004-6361/200913658},
archivePrefix = {arXiv},
       eprint = {1006.1621},
 primaryClass = {astro-ph.SR},
       adsurl = {https://ui.adsabs.harvard.edu/abs/2010A&A...520A..86Z}
}

@INPROCEEDINGS{VDH1976,
       author = {{van den Heuvel}, E.~P.~J.},
        title = "{Late Stages of Close Binary Systems}",
    booktitle = {Structure and Evolution of Close Binary Systems},
         year = 1976,
       editor = {{Eggleton}, Peter and {Mitton}, Simon and {Whelan}, John},
       series = {IAU Symposium},
       volume = {73},
        month = jan,
        pages = {35},
       adsurl = {https://ui.adsabs.harvard.edu/abs/1976IAUS...73...35V}
}

@ARTICLE{Ruderman1971,
       author = {{Ruderman}, M.~A. and {Spiegel}, E.~A.},
        title = "{Galactic Wakes}",
      journal = {\apj},
         year = 1971,
        month = apr,
       volume = {165},
        pages = {1},
          doi = {10.1086/150870},
       adsurl = {https://ui.adsabs.harvard.edu/abs/1971ApJ...165....1R}
}

@ARTICLE{Andronov1990,
       author = {{Andronov}, I.~L. and {Yavorskij}, Yu. B.},
        title = "{On the moments of inertia and radii of the white dwarfs and polytropic stars}",
      journal = {Contributions of the Astronomical Observatory Skalnate Pleso},
         year = 1990,
        month = jan,
       volume = {20},
        pages = {155},
       adsurl = {https://ui.adsabs.harvard.edu/abs/1990CoSka..20..155A}
}

@ARTICLE{Dokuchaev1964,
       author = {{Dokuchaev}, V.~P.},
        title = "{Emission of Magnetoacoustic Waves in the Motion of Stars in Cosmic Space.}",
      journal = {\sovast},
         year = 1964,
        month = aug,
       volume = {8},
        pages = {23},
       adsurl = {https://ui.adsabs.harvard.edu/abs/1964SvA.....8...23D}
}

@ARTICLE{Shadmehri2012,
       author = {{Shadmehri}, Mohsen and {Khajenabi}, Fazeleh},
        title = "{Dynamical friction in a magnetized gas}",
      journal = {\mnras},
         year = 2012,
        month = aug,
       volume = {424},
       number = {2},
        pages = {919-926},
          doi = {10.1111/j.1365-2966.2012.21237.x},
archivePrefix = {arXiv},
       eprint = {1205.0795},
 primaryClass = {astro-ph.GA},
       adsurl = {https://ui.adsabs.harvard.edu/abs/2012MNRAS.424..919S}
}

@ARTICLE{Sanchez2012,
       author = {{S{\'a}nchez-Salcedo}, F.~J.},
        title = "{Dynamical Friction in a Gaseous Medium with a Large-scale Magnetic Field}",
      journal = {\apj},
         year = 2012,
        month = feb,
       volume = {745},
       number = {2},
          eid = {135},
        pages = {135},
          doi = {10.1088/0004-637X/745/2/135},
archivePrefix = {arXiv},
       eprint = {1111.6632},
 primaryClass = {astro-ph.CO},
       adsurl = {https://ui.adsabs.harvard.edu/abs/2012ApJ...745..135S}
}

@article{Desjacques2022,
doi = {10.3847/1538-4357/ac5519},
url = {https://dx.doi.org/10.3847/1538-4357/ac5519},
year = {2022},
month = {mar},
publisher = {The American Astronomical Society},
volume = {928},
number = {1},
pages = {64},
author = {Desjacques, Vincent and Nusser, Adi and Bühler, Robin},
title = {Analytic Solution to the Dynamical Friction Acting on Circularly Moving Perturbers},
journal = {The Astrophysical Journal}
}

@ARTICLE{kim2010,
       author = {{Kim}, Woong-Tae},
        title = "{Nonlinear Dynamical Friction of a Circular-orbit Perturber in a Gaseous Medium}",
      journal = {\apj},
         year = 2010,
        month = dec,
       volume = {725},
       number = {1},
        pages = {1069-1081},
          doi = {10.1088/0004-637X/725/1/1069},
archivePrefix = {arXiv},
       eprint = {1010.1995},
 primaryClass = {astro-ph.GA},
       adsurl = {https://ui.adsabs.harvard.edu/abs/2010ApJ...725.1069K}
}

@ARTICLE{kim2008,
       author = {{Kim}, Hyosun and {Kim}, Woong-Tae and {S{\'a}nchez-Salcedo}, F.~J.},
        title = "{Dynamical Friction of Double Perturbers in a Gaseous Medium}",
      journal = {\apjl},
         year = 2008,
        month = may,
       volume = {679},
       number = {1},
        pages = {L33},
          doi = {10.1086/589149},
archivePrefix = {arXiv},
       eprint = {0804.2010},
 primaryClass = {astro-ph},
       adsurl = {https://ui.adsabs.harvard.edu/abs/2008ApJ...679L..33K}
}

@ARTICLE{kim2007,
       author = {{Kim}, Hyosun and {Kim}, Woong-Tae},
        title = "{Dynamical Friction of a Circular-Orbit Perturber in a Gaseous Medium}",
      journal = {\apj},
         year = 2007,
        month = aug,
       volume = {665},
       number = {1},
        pages = {432-444},
          doi = {10.1086/519302},
archivePrefix = {arXiv},
       eprint = {0705.0084},
 primaryClass = {astro-ph},
       adsurl = {https://ui.adsabs.harvard.edu/abs/2007ApJ...665..432K}
}

@article{kim2009,
doi = {10.1088/0004-637X/703/2/1278},
url = {https://dx.doi.org/10.1088/0004-637X/703/2/1278},
year = {2009},
month = {sep},
publisher = {The American Astronomical Society},
volume = {703},
number = {2},
pages = {1278},
author = {Kim, Hyosun and Kim, Woong-Tae},
title = {NONLINEAR DYNAMICAL FRICTION IN A GASEOUS MEDIUM},
journal = {The Astrophysical Journal}
}

@ARTICLE{Vetter2024,
       author = {{Vetter}, Marco and {R{\"o}pke}, Friedrich K. and {Schneider}, Fabian R.~N. and {Pakmor}, R{\"u}diger and {Ohlmann}, Sebastian T. and {Lau}, Mike Y.~M. and {Andrassy}, Robert},
        title = "{From spherical stars to disk-like structures: 3D common-envelope evolution of massive binaries beyond inspiral}",
      journal = {\aap},
         year = 2024,
        month = nov,
       volume = {691},
          eid = {A244},
        pages = {A244},
          doi = {10.1051/0004-6361/202451579},
archivePrefix = {arXiv},
       eprint = {2410.07841},
 primaryClass = {astro-ph.SR},
       adsurl = {https://ui.adsabs.harvard.edu/abs/2024A&A...691A.244V}
}

@ARTICLE{Vetter2025,
       author = {{Vetter}, Marco and {R{\"o}pke}, Friedrich K. and {Schneider}, Fabian R.~N. and {Pakmor}, R{\"u}diger and {Ohlmann}, Sebastian and {Mor{\'a}n-Fraile}, Javier and {Lau}, Mike Y.~M. and {Leidi}, Giovanni and {Gagnier}, Damien and {Andrassy}, Robert},
        title = "{Magnetically driven outflows in the 3D common envelope evolution of massive stars}",
      journal = {\aap},
         year = 2025,
        month = jun,
       volume = {698},
          eid = {A133},
        pages = {A133},
          doi = {10.1051/0004-6361/202554685},
archivePrefix = {arXiv},
       eprint = {2504.12213},
 primaryClass = {astro-ph.SR},
       adsurl = {https://ui.adsabs.harvard.edu/abs/2025A&A...698A.133V}
}

@ARTICLE{MM2017,
       author = {{MacLeod}, Morgan and {Antoni}, Andrea and {Murguia-Berthier}, Ariadna and {Macias}, Phillip and {Ramirez-Ruiz}, Enrico},
        title = "{Common Envelope Wind Tunnel: Coefficients of Drag and Accretion in a Simplified Context for Studying Flows around Objects Embedded within Stellar Envelopes}",
      journal = {\apj},
         year = 2017,
        month = mar,
       volume = {838},
       number = {1},
          eid = {56},
        pages = {56},
          doi = {10.3847/1538-4357/aa6117},
archivePrefix = {arXiv},
       eprint = {1704.02372},
 primaryClass = {astro-ph.SR},
       adsurl = {https://ui.adsabs.harvard.edu/abs/2017ApJ...838...56M}
}

@ARTICLE{Blackman2014,
       author = {{Blackman}, E.~G. and {Lucchini}, S.},
        title = "{Using kinematic properties of pre-planetary nebulae to constrain engine paradigms.}",
      journal = {\mnras},
         year = 2014,
        month = may,
       volume = {440},
        pages = {L16-L20},
          doi = {10.1093/mnrasl/slu001},
archivePrefix = {arXiv},
       eprint = {1312.5372},
 primaryClass = {astro-ph.SR},
       adsurl = {https://ui.adsabs.harvard.edu/abs/2014MNRAS.440L..16B}
}

@article{AA,
   author    = {{Fejes I.}},
   title     = {SS 433 extended radio structure observed with the European VLBI network},
   journal   = AA,
   year      = 1986,
   volume    = 168,
   pages     = 69 
}

@article{Gottlieb2009,
	author = {Gottlieb, Sigal and Ketcheson, David I. and Shu, Chi-Wang},
	date = {2009/03/01},
	doi = {10.1007/s10915-008-9239-z},
	id = {Gottlieb2009},
	isbn = {1573-7691},
	journal = {Journal of Scientific Computing},
	number = {3},
	pages = {251--289},
	title = {High Order Strong Stability Preserving Time Discretizations},
	url = {https://doi.org/10.1007/s10915-008-9239-z},
	volume = {38},
	year = {2009}}

@article{Liu2023,
doi = {10.1088/1674-4527/acd89e},
url = {https://doi.org/10.1088/1674-4527/acd89e},
year = {2023},
month = {jul},
publisher = {National Astromonical Observatories, CAS and IOP Publishing},
volume = {23},
number = {8},
pages = {082001},
author = {Liu, Zheng-Wei and Röpke, Friedrich K. and Han, Zhanwen},
title = {Type Ia Supernova Explosions in Binary Systems: A Review},
journal = {Research in Astronomy and Astrophysics}
}

@ARTICLE{Li2024,
       author = {{Li}, Zhenwei and {Chen}, Xuefei},
        title = "{Compact objects in close orbits as gravitational wave sources: Formation scenarios and properties}",
      journal = {Results in Physics},
         year = 2024,
        month = apr,
       volume = {59},
          eid = {107568},
        pages = {107568},
          doi = {10.1016/j.rinp.2024.107568},
archivePrefix = {arXiv},
       eprint = {2403.06358},
 primaryClass = {astro-ph.SR},
       adsurl = {https://ui.adsabs.harvard.edu/abs/2024ResPh..5907568L}
}

@ARTICLE{Wei2024,
       author = {{Wei}, Dandan and {Schneider}, Fabian R.~N. and {Podsiadlowski}, Philipp and {Laplace}, Eva and {R{\"o}pke}, Friedrich K. and {Vetter}, Marco},
        title = "{Evolution and final fate of massive post-common-envelope binaries}",
      journal = {\aap},
         year = 2024,
        month = aug,
       volume = {688},
          eid = {A87},
        pages = {A87},
          doi = {10.1051/0004-6361/202348560},
archivePrefix = {arXiv},
       eprint = {2311.07278},
 primaryClass = {astro-ph.HE},
       adsurl = {https://ui.adsabs.harvard.edu/abs/2024A&A...688A..87W}
}

@ARTICLE{Thun2016,
       author = {{Thun}, Daniel and {Kuiper}, Rolf and {Schmidt}, Franziska and {Kley}, Wilhelm},
        title = "{Dynamical friction for supersonic motion in a homogeneous gaseous medium}",
      journal = {\aap},
         year = 2016,
        month = may,
       volume = {589},
          eid = {A10},
        pages = {A10},
          doi = {10.1051/0004-6361/201527629},
archivePrefix = {arXiv},
       eprint = {1601.07799},
 primaryClass = {astro-ph.GA},
       adsurl = {https://ui.adsabs.harvard.edu/abs/2016A&A...589A..10T}
}

\begin{appendix}
\section{Extra figures and tables}\label{app:Extra_figures}

\begin{table}[htbp]
\centering
\caption{Polynomial coefficients $a_m$ and $b_m$.}
\label{tab:fit}
\begin{tabular}{c c c c c c}
\hline
$a_0$ & $a_1$ & $a_2$ & $a_3$& $b_0$  \\
\hline
-0.455394 & 0.310163 & -0.0111443 & -3.21466  & -0.0737447 \\
\hline
\end{tabular}
\end{table}
\begin{figure}[h]
\resizebox{\hsize}{!}{\includegraphics{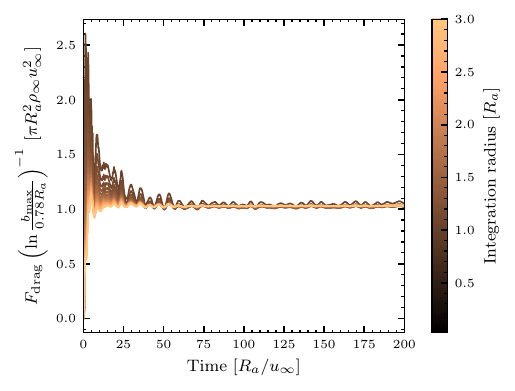}}
\caption{Time evolution of the gravitational drag measured within different logarithmically spaced integration radii, scaled by the Coulomb logarithm. We find the gravitational drag to follow the Coulomb-logarithm scaling in a quasi-steady state, provided $b_{\rm min} \simeq 0.78$.}
\label{fig:coulomb}
\end{figure}
\begin{figure}
\resizebox{\hsize}{!}{\includegraphics{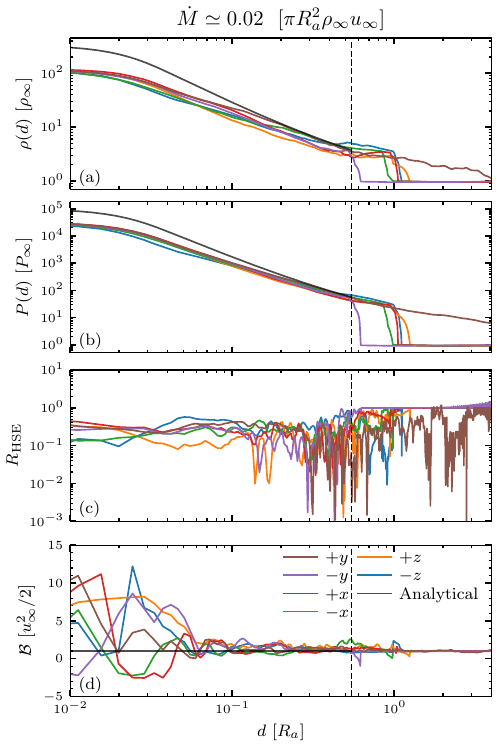}}
 \caption{Same as Fig.~\ref{fig:HSE_bubble} but for accretion parameters $\delta = 0$ and $\gamma = 100$, corresponding to a mass accretion $\dot{M} \simeq 0.02\, \dot{M}_{\rm  HL}$ in a quasi-steady state.}
    \label{fig:HSE_bubble_acc}
\end{figure}
 \begin{figure}[]
\resizebox{\hsize}{!}{\includegraphics{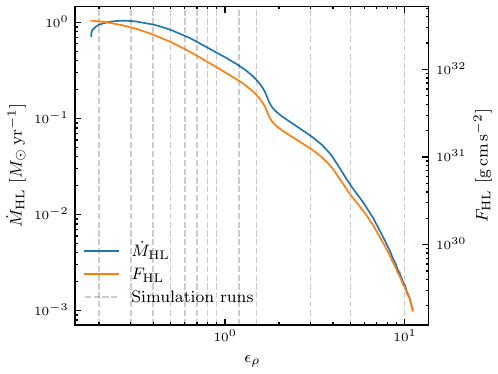}}
    \caption{Hoyle--Lyttleton accretion rate and drag force as a function of the stratification parameter $(\epsilon_\rho$) for the $2M_\odot$ red giant MESA model (see Fig.~\ref{fig:Prhoini}). Vertical dashed lines indicate the $\epsilon_\rho$ values employed in the stratified simulations of this work.}
    \label{fig:MHL_prof}
\end{figure}
 \begin{figure}[]
\resizebox{\hsize}{!}{\includegraphics{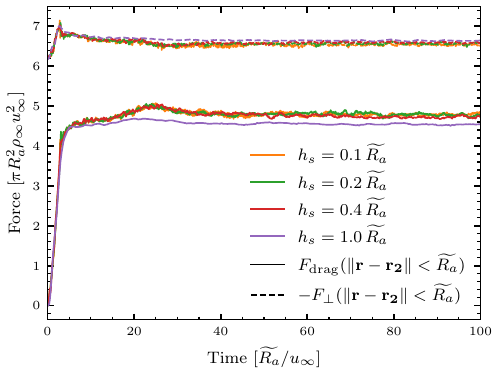}}
    \caption{Time evolution of the radial and drag forces exerted by the gas on the companion in stratified simulations with $\epsilon_\rho= 3$, with different softening radii ($h_s$). Forces are integrated within a sphere of radius $3\, \widetilde{R_a}$.}
    \label{fig:Drag_rsoft}
\end{figure}
\begin{figure}[]
\resizebox{\hsize}{!}{\includegraphics{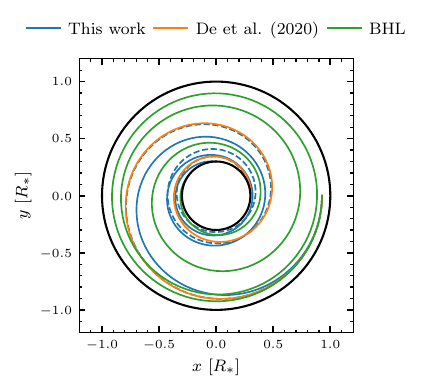}}
\caption{Inspiral of a $0.2\,M_\odot$ companion through the envelope of a $2\,M_\odot$ red giant, using four drag force prescriptions: our fitted drag and lift prescriptions (Eqs.~\ref{eq:fit_fdrag}-\ref{eq:fit_fperp}; solid blue); the prescription from \citet[orange]{De2020}; the Hoyle--Lyttleton formula, $F_{\rm drag} = \pi R_a^2 \rho_\infty u_\infty^2$ (green); and our fitted drag prescription but with $F_\perp$ set to zero (dashed blue). For clarity, trajectories are shown only for $r \ge 0.25$. Below this radius, trajectories start intersecting with themselves, corresponding to interaction with the companion's own wake, which is not included in our model. At such small orbital separations, interactions with the primary's wake would also play a major role in the orbital evolution. }
\label{fig:orbit}
\end{figure}
\begin{figure*}
    \centering
    \resizebox{0.99\hsize}{!}{\includegraphics{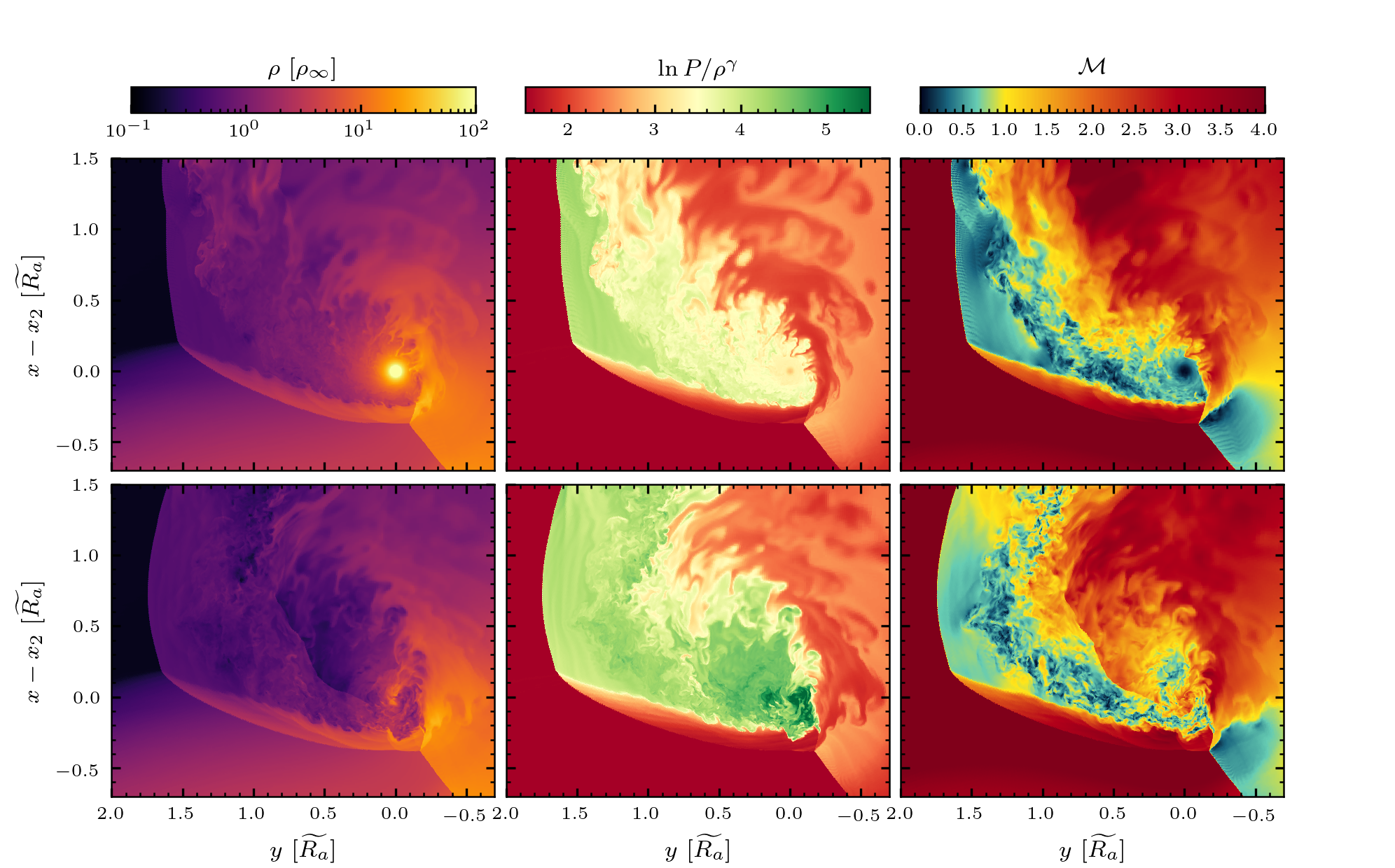}}
    \caption{Zoomed-in snapshots in the $xy$ plane of the density, pseudo-entropy, and Mach number at $t = 75\, \widetilde{R_a}/u_\infty$ for stratified simulations with $h_s = 0.05\, \widetilde{R_a}$ and $\epsilon_\rho = 3$. First row: Non-accreting case. Second row: Accreting case with $\gamma = 1000$ and $\delta = 1$.}
        \label{fig:snap_acc_strat_xy}
\end{figure*}
\begin{figure*}[]
    \resizebox{0.99\hsize}{!}{\includegraphics{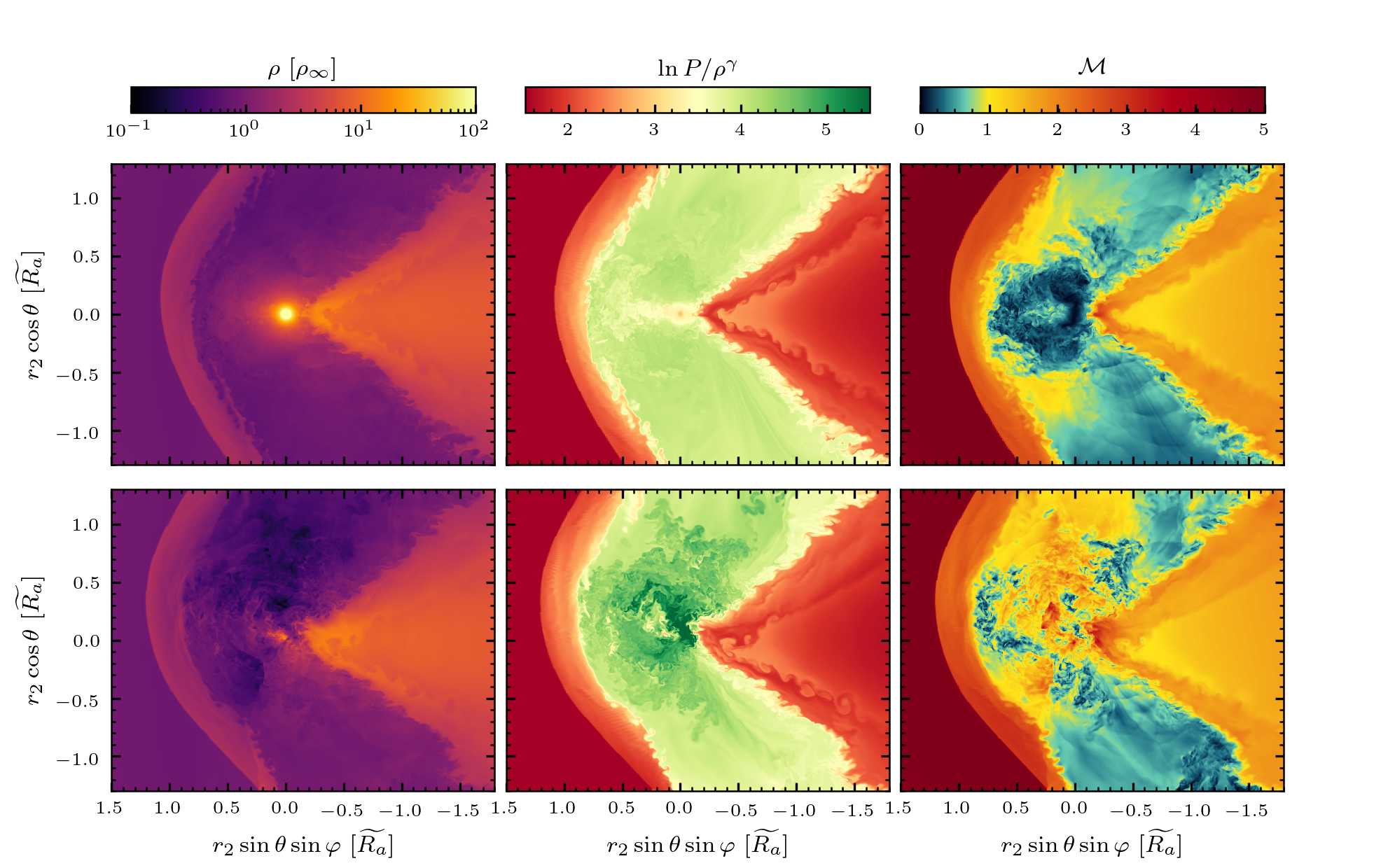}}
    \caption{Same as Fig.~\ref{fig:snap_acc_strat_xy} but on a spherical shell at radius $r_2$.}
        \label{fig:snap_acc_strat_xz}
\end{figure*}
\begin{figure*}[]
\sidecaption
    \includegraphics[width=12cm]{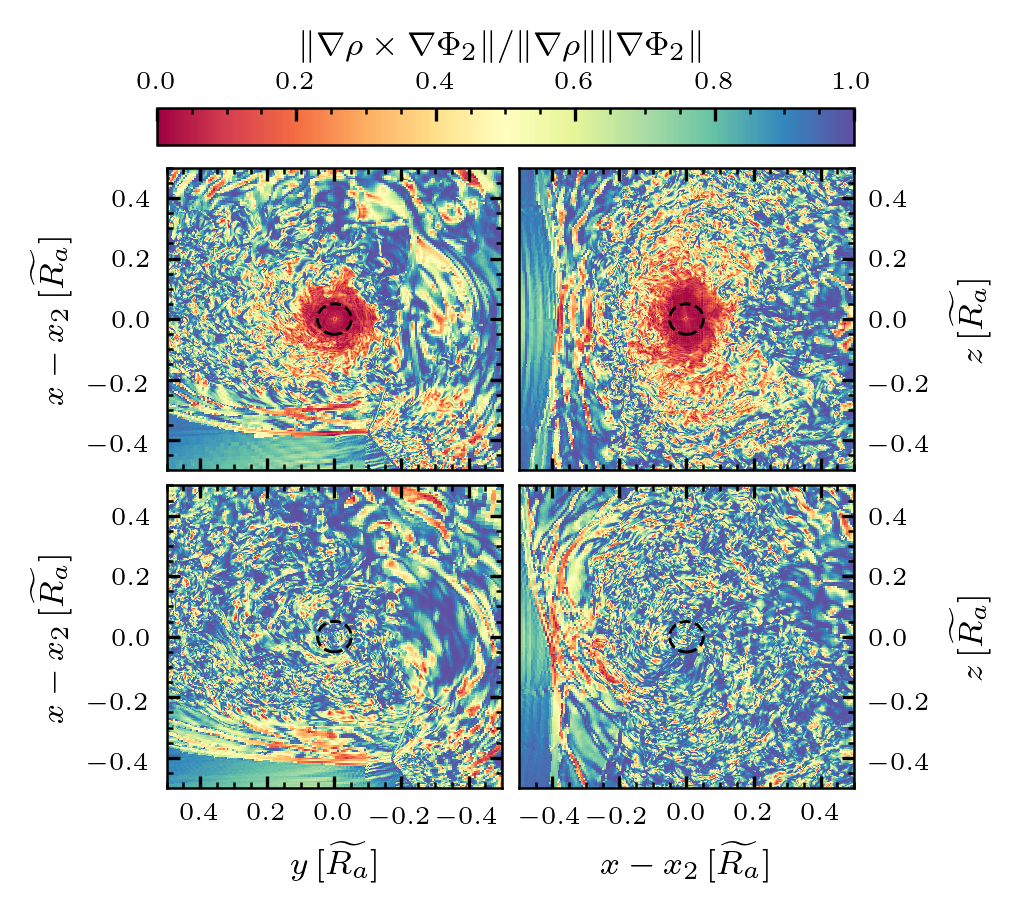}
\caption{Close-up view of the normalized deviation from hydrostatic equilibrium in $xy$ (left column) and $xz$ (right column) planes at $t = 100\, \widetilde{R_a}/u_\infty$ for two simulations with $\epsilon_\rho = 3$. Top row: Case without accretion. Bottom row: Case with accretion with $\delta = 0$ and $\gamma = 100$. Black circles indicate the softening radius $h_s = 0.05\, \widetilde{R_a}$.}
\label{fig:HE}
\end{figure*}

\begin{table*}[!tbp]
\centering
\begin{threeparttable}
\caption{Section~\ref{sec:strat}'s simulation parameters and results.}
\label{tab:rotstratacc}
\begin{tabular}{c c c c c c c c c c}
\hline
$\epsilon_\rho$ & $\gamma$ & $\delta$ & $ \langle F_{\rm drag} \rangle $ & $\langle -\dot{p}_\varphi \rangle/ \langle F_{\rm drag}  \rangle $ & $\langle F_\perp \rangle $ &  $\langle \dot{p}_r \rangle / \langle F_\perp  \rangle $ & $\langle \dot{M} \rangle$ & $\langle \ell_z/\ell_{\rm K} \rangle$ & $\langle \tau_{\rm spin-up} \rangle$\\
\hline
3 & 0  & -- & 4.85 & -- & $-6.53$ & -- &  -- & -- & -- \\
3 & 1000 & 1 & 4.81 & $-2.32\cdot 10^{-3}$ & $-6.67$ & $3.16 \cdot 10^{-3}$ & $1.38 \cdot 10^{-2}$ & $0.81$ & $273.99$\\
0.5 & 0 & -- & 0.77 & -- & $0.32$ & -- & -- & -- & --\\
$0.5$ & 1000 & 1 & 0.77 & $-5.57 \cdot 10^{-2}$ & $3.56 \cdot 10^{-2}$ & $-0.83$ & $4.39 \cdot 10^{-2}$ & 0.26 & $49.30$\\
0.3 & 0 & -- & 0.40 & -- & $0.27$ & -- & -- & -- & --\\
$0.3$ & 1000 & 1 & $0.37$ & $-7.25 \cdot 10^{-2}$ & $0.19$ & $-0.21$ & $3.09 \cdot 10^{-2}$  & $5.23 \cdot 10^{-2}$ & $188.30$\\
\hline
\end{tabular}
\tablefoot{Brackets indicate a time average over $t \in [50,100]\, \widetilde{R_a}/u_\infty$. Drag and lift forces, as well as linear momentum accretion rates are expressed in units $F_{\rm HL}$. The spin-up timescale in expressed in orbital periods. In the inertial frame, the companion orbits in the $+\be_\varphi$ direction, $-\dot{p}_\varphi$ thus effectively acts as a drag force.} 

\end{threeparttable}
\end{table*}

\FloatBarrier 
\section{Simulation stochasticity}\label{app:stat}
We performed 15 separate simulations, each initialized with a random perturbation 
$\delta \epsilon_\rho \in 0.5 \times [-10^{-3},\,10^{-3}]$ to the background stratification parameter $\epsilon_\rho = 0.5$, 
in order to capture the system's intrinsic stochastic behavior. 
This ensemble allows us to assess how much of the observed variability arises from inherent randomness, 
as opposed to variations in global parameters or added physical mechanisms such as magnetic fields, rotation, or accretion.

\begin{figure}
    \centering    \includegraphics[width=0.5\textwidth]{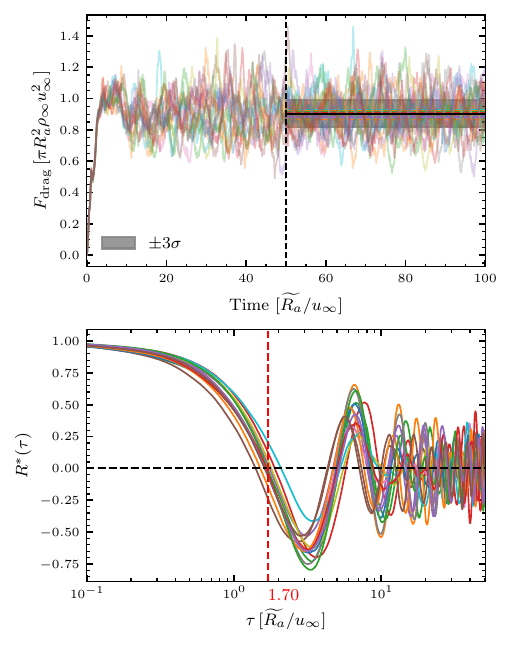}
    \caption{Top: Time evolution of the drag force exerted by the gas on the companion in 15 stratified simulations with 
$\epsilon_\rho = 0.5 + \delta \epsilon_\rho$. Forces are integrated within a sphere of radius $3\, \widetilde{R_a}$. 
The shaded region shows the $3\sigma$ confidence interval. Bottom: Autocorrelation function of the detrended drag force. The vertical red line marks the mean de-correlation time.}
    \label{fig:A}
\end{figure}

To quantify this stochasticity, we computed the autocorrelation function of the 
detrended drag force in each simulation run. 
We began with the raw drag force,  $F_{\rm drag} \equiv F_{\rm drag}(\|\br - \br_2\| < 3\widetilde{R_a})$, 
and subtracted its best-fit linear trend to suppress spurious long-term correlations. 
To further reduce edge effects in the autocorrelation calculation, we mirrored the detrended signal by appending its time-reversed copy, 
following the method of \citet{Andrassy2022}. The resulting time series, $F^\ast_{\rm drag}(t),$ was then used to compute the autocorrelation function over the interval $t \in [t_0,\,t_{\rm end}]\,\widetilde{R_a}/u_\infty$:
\begin{equation}
R^\ast(\tau) = \frac{\int_{t_0}^{t_{\rm end}} F^\ast_{\rm drag}(t) F^\ast_{\rm drag}(t + \tau) \,\mathrm{d}t}
{\int_{t_0}^{t_{\rm end}} F^\ast_{\rm drag}(t)^2 \,\mathrm{d}t} \,,
\end{equation}
where $t_0 = 50$ and $t_{\rm end} = 100$. 
The de-correlation time $\tau_{\rm dec}$ is defined as the lag $\tau$ at which the normalized autocorrelation function $R^\ast(\tau)$ first crosses zero. This provides a characteristic timescale over which fluctuations in the time series become approximately independent. For each realization, we estimated the number of independent fluctuations as
\begin{equation}
N_{\rm real} = \frac{t_{\rm end} - t_0}{\tau_{\rm dec}} \,.
\end{equation}
For each realization, we computed the standard deviation $\sigma_0$ of the original (not detrended) drag force $F_{\rm drag}$ and estimate the standard error as
\begin{equation}
\sigma_i = \frac{\sigma_{0}}{\sqrt{N_{{\rm eff}}}} \ .
\end{equation}
Figure~\ref{fig:A} shows the time evolution of the drag force for all 15 simulations, including their individual time-averaged values over the interval $t \in [t_0,,t_{\rm end}],\widetilde{R_a}/u_\infty$, the standard error averaged across the realizations, and the autocorrelation function $R(\tau)$. We find that, for a softening radius $h_s = 0.1\, \widetilde{R_a}$, a stratification parameter $\epsilon_\rho = 0.5 = \delta \epsilon_\rho$, and in the absence of accretion, the drag force exhibits significant time variability within each realization. However, the time-averaged drag forces differ by at most $\sim 4\%$ across the 15 runs, with a 3$\sigma$ range corresponding to a maximum relative variation of $\sim 17\%$. This range quantifies the baseline stochastic variability arising from small random perturbations in the stratification. The 3$\sigma$ range observed here provides an illustration of the typical amplitude of fluctuations due to intrinsic stochasticity; their behavior under other physical conditions remains uncertain, however.

\end{appendix}
\end{document}